\documentclass[aps,prb,floatfix,superscriptaddress,twocolumn,longbibliography]{revtex4-2}

\usepackage{graphicx}
\usepackage{diagbox}
\usepackage{amsmath,bm}
\usepackage{array}    

\usepackage{subcaption} 
\usepackage{amssymb}
\usepackage{booktabs}
\usepackage{braket}
\usepackage{amsfonts}
\usepackage{placeins}
\usepackage{dsfont}
\usepackage{comment}
\usepackage{xcolor}
\usepackage{lipsum}
\usepackage{hyperref}
\usepackage{pifont}

\usepackage{float}
\usepackage{dblfloatfix}
\usepackage{placeins}
\usepackage{bbm}

\usepackage{amsmath}    

\usepackage[justification=justified]{ragged2e}
\usepackage{etoolbox}
\makeatletter
\patchcmd{\@makecaption}
  {\centering}
  {\justifying}
  {}
  {}
\makeatother
\makeatletter
\long\def\@makecaption#1#2{%
  \vskip\abovecaptionskip
  \begingroup
    \justifying
    \sbox\@tempboxa{#1: #2}%
    #1: #2\par
  \endgroup
  \vskip\belowcaptionskip
}

\hypersetup{colorlinks=true,linkcolor=blue,anchorcolor=blue,citecolor=blue,filecolor=blue,urlcolor=blue,bookmarksnumbered=true,pdfview=FitB}

\begin{document}

\title{Phase-controlled perfect nonlocal spin and charge diode effects in a four-terminal Josephson junction with $p$-wave magnets}

\author{Lovy Sharma}
\affiliation{Department of Physics, Indian Institute of Technology Delhi, Hauz Khas, New Delhi, India 110016}
\author{Bimal Ghimire}
\affiliation{Department of Physics, Indian Institute of Technology Hyderabad, Kandi, Sangareddy, Telangana, India 502285}
\author{Manisha Thakurathi}
\affiliation{Department of Physics, Indian Institute of Technology Hyderabad, Kandi, Sangareddy, Telangana, India 502285}
\date{\today}

\begin{abstract}
We theoretically investigate charge and spin transport in a four-terminal Josephson junction with a normal-metal barrier.
The top and bottom superconducting leads are equal-spin triplet $p_{y}$-wave superconductors, while the left and right leads are $p$-wave magnets with proximity-induced conventional $s$-wave superconductivity. When the transverse macroscopic phase difference between the top and bottom leads is set to zero, a longitudinal phase bias generates a pure transverse spin current with perfect $100\%$ nonreciprocity. Remarkably, a finite transverse phase difference preserves the perfect spin-diode effect while simultaneously inducing a perfect charge-diode effect, enabling fully nonreciprocal spin and charge transport. Moreover, the spin-diode efficiency exhibits sharp, step-like switching as a function of both the gate voltage applied to the barrier and the crystallographic orientation of the $p$-wave magnet, providing independent and experimentally accessible knobs for controlling the diode polarity. The diode response remains robust against asymmetric interface couplings, nonmagnetic disorder, variations in the relative singlet and triplet pairing strengths, temperature, and junction dimensions, demonstrating that the effect is not a consequence of fine-tuned parameters. These findings establish the proposed four-terminal junction as a highly tunable and structurally robust platform for dissipationless, phase-controlled spin and charge rectification, with potential applications in superconducting spintronics.
\end{abstract}

\maketitle

\section{Introduction}
The phenomenon of non-reciprocal charge transport under bias reversal is a cornerstone of modern solid-state electronics. From simple rectifiers to sophisticated integrated circuits, the ability to control the direction of charge flow is the backbone of modern technology \cite{6372252,6773080}. In recent years, this concept of nonreciprocity has been extended from dissipative charge transport to dissipationless supercurrents, leading to the discovery of the superconducting charge-diode effect (CDE) in both bulk superconductors (SCs) [\onlinecite{Ando2020,PhysRevB.49.9244,Miyasaka_2021,PhysRevB.107.224518,Zhang2020,Pal2022}] and Josephson junctions (JJs) [\onlinecite{Baumgartner2022,PhysRevLett.131.196301,PhysRevX.12.041013,PhysRevB.106.214524,PhysRevLett.131.096001,PhysRevLett.129.267702}]. The superconducting CDE manifests itself as an asymmetry between the critical currents for opposite bias directions, thus the system remains dissipationless when current flows one way, while turns resistive when it is reversed. In JJs, this directional asymmetry originates from the phase-dependent Josephson current, where opposite current directions exhibit unequal critical currents under a superconducting phase bias. 

In the latest developments, the concept of superconducting nonreciprocity has been generalized from charge transport to the transport of spin angular momentum, giving rise to the superconducting spin-diode effect (SDE). In the superconducting SDE, the critical spin supercurrents flowing in opposite directions become unequal, enabling nonreciprocal spin transport without energy dissipation. The superconducting SDE has been theoretically proposed in a wide range of hybrid structures, including systems with non-coplanar magnetic textures and spin-active interfaces [\onlinecite{nb38-v1jq,4t18-yyx4,nikolic2025necessaryconditionsspinresolvedjosephson}], as well as ferromagnetic multilayers [\onlinecite{h2qg-qhf7}]. In parallel, a general microscopic description of the superconducting SDE driven by SOC has been established within the Ginzburg-Landau framework [\onlinecite{PhysRevLett.132.216001}]. A spin supercurrent generally requires spin-triplet superconducting correlations. Such triplet correlations may arise intrinsically in unconventional spin-triplet SCs or be induced via the superconducting proximity effect in conventional spin-singlet SCs. The latter mechanism relies on spin-dependent scattering and symmetry breaking at superconducting interfaces, which can be engineered through spin-active interfaces, noncollinear magnetization, magnetic textures, or the interplay between SOC and magnetic exchange fields [\onlinecite{Bao2013,PhysRevB.105.184511,patra2026floquetmajoranaflatbands}]. Due to its ability to rectify spin transport without dissipation, the superconducting SDE provides a promising platform for superconducting spintronics, with potential applications in non-reciprocal spin-current, energy-efficient spin-based logic, and ultrasensitive magnetic sensing [\onlinecite{RevModPhys.80.1517,RevModPhys.76.323,Linder2015,Eschrig_2015}].

From a symmetry perspective, in general, the superconducting CDE requires the simultaneous breaking of time-reversal symmetry (TRS) and inversion symmetry (IS), whereas the SSDE only necessitates the breaking of IS \cite{yqsg-xdg8,PhysRevB.106.214524,PhysRevB.110.014518,sharma2026pwavemagnetdrivenfieldfree,PhysRevLett.132.216001}. In conventional superconducting hybrid structures, these symmetry requirements are typically realized through spin-active interfaces, Rashba spin-orbit coupling (SOC), noncoplanar magnetic textures, external magnetic fields, or ferromagnetic layers \cite{h2qg-qhf7,PhysRevB.106.214524,PhysRevLett.132.216001,4t18-yyx4,nb38-v1jq}. However, reliance on external magnetic fields or ferromagnets introduces stray magnetic fields that are detrimental to superconducting coherence and present a major obstacle to scalable superconducting device architectures. To circumvent these limitations, recent theoretical studies have increasingly focused on unconventional magnetic phases as an alternative route to realizing superconducting diode effects [\onlinecite{fu2026perfectspinnonreciprocitygated,PhysRevB.110.014518,yqsg-xdg8,sharma2026pwavemagnetdrivenfieldfree,Yang_2026,Wu2022,Lin2022,Sibgat,debnath2024,zhao2026spinpolarizedjosephsoncurrentinduced,pal2026emergentsuperconductingphasesunconventional,g4ry-j1xy}]. These unconventional magnets have emerged from a symmetry-based classification that extends beyond conventional magnetic space groups by incorporating spin-group symmetry, which allows spin and real-space operations to transform independently in the non-relativistic limit [\onlinecite{PhysRevX.12.040002,PhysRevX.12.040501,PhysRevX.12.031042,PhysRevX.12.011028,hellenes2024pwavemagnets}]. This framework has uncovered novel magnetic phases, including altermagnets (AMs), characterized by even-parity momentum-dependent spin splitting [\onlinecite{PhysRevX.12.031042}], and $p$-wave magnets (PMs), which exhibit odd-parity spin splitting [\onlinecite{hellenes2024pwavemagnets}]. The absence of net magnetization, together with anisotropic spin splitting, makes these materials particularly attractive for superconducting circuits, where eliminating the stray magnetic fields associated with ferromagnets is crucial.

Recently, the superconducting CDE in two-terminal JJs has been extended to 
multiterminal JJs \cite{Coraiola2024,Gupta2023,Sahoo_2025,
10.21468/SciPostPhys.17.2.037,sahoo2026giantfieldfreetransversejosephson}.
In an $N$-terminal JJ, the Josephson current is governed by $N-1$ independent macroscopic phase differences, greatly enhancing the available parameter space for tailoring nonreciprocal superconducting transport \cite{Riwar2016,PhysRevB.90.155450}. In particular, four-terminal JJs have been identified as a versatile platform to realize a transverse superconducting CDE, where a longitudinal phase bias induces a nonreciprocal transverse charge supercurrent \cite{Sahoo_2025,sahoo2026giantfieldfreetransversejosephson}.
Moreover, for suitable parameter regimes, such phase-engineered multiterminal architectures can even support strongly asymmetric or nearly unidirectional Josephson transport \cite{Sahoo_2025, sahoo2026giantfieldfreetransversejosephson}. Despite these advances, the superconducting SDE in multiterminal JJs, particularly in four-terminal geometries, remains largely unexplored. This gap is especially intriguing in light of recent predictions that PMs naturally generate transverse spin supercurrents due to their anisotropic, odd-parity spin-split Fermi surface \cite{salehi2025transversespinsupercurrentpwave,Zeng_2025}. These distinctive properties make PMs an ideal platform for investigating nonreciprocal spin transport and realizing the SSDE in multiterminal superconducting devices.

Motivated by these recent advances in the generation of superconducting CDE and SDE by unconventional magnets, we propose a four-terminal JJ comprising a normal-metal (NM) as a barrier. The left and right leads consist of PM which is proximity coupled to $s$-wave SC, while the top and bottom leads are (equal-spin) spin-triplet SCs, see Fig.[\ref{fig1aa}]. In this setup, we demonstrate a pure transverse SDE with 100\% spin polarization, driven solely by a longitudinal superconducting phase bias while the transverse phase difference is fixed at zero. Remarkably, upon introduction of a finite transverse phase difference, the junction simultaneously exhibits a 100\% spin-polarized superconducting SDE and a 100\% superconducting CDE in the transverse direction. Furthermore, we show that the polarity of the spin diode can be electrically reversed by tuning the gate potential in the NM barrier, displaying an almost step-like switching behavior. An analogous polarity reversal is achieved by rotating the crystallographic lobe angle of the PM, providing an additional magnetic control knob with similar sharp switching characteristics. Finally, to assess the robustness of these phenomena, we investigate the transverse spin and charge transport in the presence of asymmetric couplings between the left-right and top-bottom superconducting leads. We find that both the SDE and CDE remain fully robust, preserving their 100\% diode efficiency over a broad range of coupling asymmetries.

The paper is organized as follows. In Sec.~\ref{sec1}, we introduce the model and Hamiltonian describing the four-terminal planar JJ. Section~\ref{sec2} presents the Keldysh Green's function formalism used to evaluate the spin and charge currents flowing through each terminal. In Sec.~\ref{sec3}, we derive the symmetry constraints governing these currents and discuss their implications for the diode effects. The numerical results and their physical interpretation are presented in Sec.~\ref{sec4}. Subsequently, Sec.~\ref{sec5} examines the robustness of the diode effect against interface asymmetry, disorder, and parameter variations, while Sec.~\ref{sec6} summarizes our main findings. The Appendices provide the matrix representations of the Hamiltonian and transformation operators, together with details of the surface and nonlocal Green's functions.

\section{Model and Hamiltonian}
\label{sec1}
\begin{figure}
    \centering
    \includegraphics[width=0.9\linewidth]{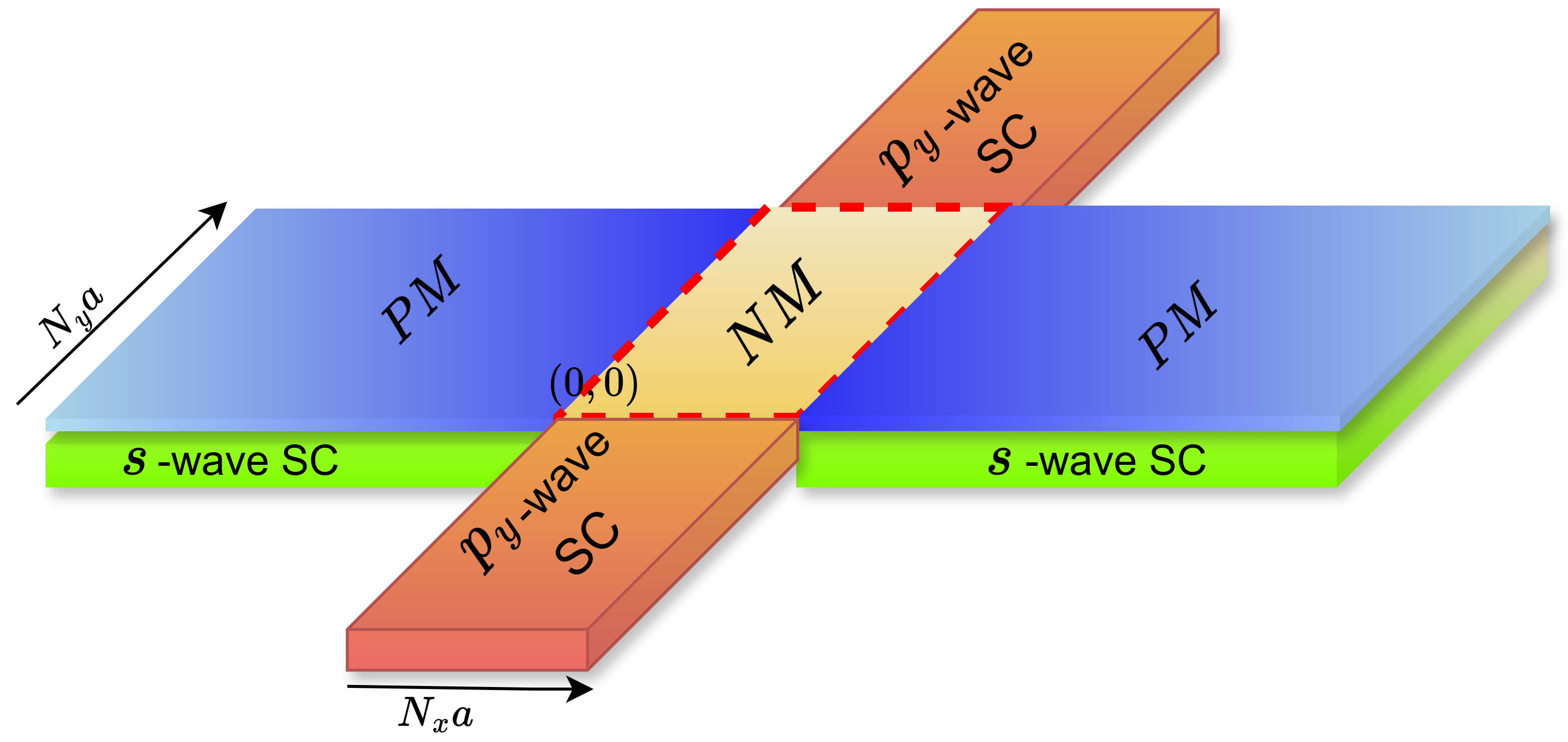}
    \caption{Schematic illustration of our four-terminal model, where the barrier region is NM coupled to four semi-infinite leads at the top, bottom, left, and right. The left and right leads host proximity-induced
$s$-wave superconductivity in PM, while the top and bottom leads are $p_y$-wave SCs.}
    \label{fig1aa}
\end{figure}
We consider a four terminal JJ in which the NM region acts as a common barrier connecting all superconducting terminals. The SC leads attached along $+x$ and $-x$ are modeled as semi-infinite PMs with proximity induced $s$-wave superconducting pairing. In contrast, the leads attached along the $+y$ and $-y$ axes are taken to be semi-infinite $p_y$-wave SC. The geometry of the proposed device is illustrated in Fig.[\ref{fig1aa}].
Such a hybrid multi-terminal configuration provides a versitile platform to study conventional spin-singlet and unconventional spin-triplet superconducting corelations mediated through NM barrier region. The full Hamiltonian describing this four terminal JJ can be decomposed into a contribution from individual SC leads, NM barrier region, and coupling between and thus can be written as
\begin{align}
    \mathcal{H}=H_{L} + H_{R} + H_{T} + H_{B} + H_{N} +H_{C} .
    \label{eq1}
\end{align}

where $H_L$, $H_R$, $H_T$, and $H_B$ denote the Hamiltonians of the left, right, top, and bottom superconducting leads, respectively, $H_N$ describes the central normal region and $H_C$ accounts for the coupling between the NM region and SC leads. Next, we write the continuum form of Hamiltonian of PM which is proximity coupled to $s$-wave SC, $p_y$-wave SC, and the NM barrier region as follows:
\begin{align}
    \nonumber
    H_{\nu}=&-2t_0(\cos{k_x} +\cos{k_y}-\mu)\tau_z\rho_0\sigma_0 \\ \nonumber&+t_j(\sin{k_x}\cos{\alpha} +\sin{k_{y}}\sin{\alpha})\tau_0\rho_0\sigma_z+J_{sd}\tau_z\rho_{z}\sigma_x \\
    & -  \Delta_s(\cos{\phi_\nu}\tau_y  +   \sin{\phi_\nu}\tau_x)\rho_0 \sigma_y,
    \label{eq22}
\end{align}
\begin{align}
\nonumber
    H_{\mu}=&-2t_0(\cos{k_x} +\cos{k_y}-\mu)\tau_z\sigma_0  \\
    & -  \Delta_p\sin{k_y}(\cos{\phi_\mu}\tau_x   -   \sin{\phi_\mu}\tau_y)\sigma_z
\end{align}
\begin{align}
    H_{N}=&-2t_0(\cos{k_x} +\cos{k_y}-\mu + V_G)\tau_z\sigma_0
    \label{eq44}
\end{align}

Here, $\nu\in\{L,R\}$ and $\mu\in\{B,T\}$ denote the horizontal and vertical superconducting terminals, respectively. The superconducting phases are parameterized as $\phi_L=\phi_l/2$, $\phi_R=-\phi_l/2$, $\phi_B=\phi_t/2$, and $\phi_T=-\phi_t/2$, where $\phi_l$ ($\phi_t$) denotes the superconducting phase difference in between the left and right (top and bottom) terminals. The Pauli matrices $\tau_i$, $\rho_i$, and $\sigma_i$, with $i\in\{0,x,y,z\}$, act on the particle-hole, sectoral, and spin degrees of freedom, respectively. To describe the PM we adopt the minimal model of Ref. [\onlinecite{PhysRevLett.133.236703}]. The $p$-wave spin splitting is introduced through spin-dependent hopping $t_j$, and the net $sd$-exchange coupling of the itinerant electrons within a sector is given by  $J_{sd}$ [\onlinecite{PhysRevLett.133.236703,patra2026floquetmajoranaflatbands}]. For simplicity, we use the same spin-independent hopping $t_0$ and chemical potential $\mu$ across all regions. The pairing amplitudes of the $s$-wave and $p_y$-wave SCs are denoted by $\Delta_s$ and $\Delta_p$, respectively. We also introduce a gate potential $V_G$ in the normal barrier region. In the four-terminal JJ, the NM barrier and the $p_y$-wave superconducting leads are assumed to host no localized magnetic moments. Consequently, they do not distinguish between the two sectoral degrees of freedom, and all corresponding Hamiltonian terms are proportional to the identity matrix, $\rho_0$, in the sectoral space. The interface coupling is therefore also diagonal in the sectoral basis. As a result, the full BdG Hamiltonian of the junction is block diagonal in sectoral space, allowing the problem to decompose into two independent sectors. These sectors differ solely in the sign of the $sd$-exchange coupling, $J_{sd}$, within the PM leads. The total charge and spin currents are then obtained by summing the contributions from the two sectors. Therefore, the full Hamiltonian written in Eq.~(\ref{eq1})  also takes a block diagonal in sectorial basis, given by 
\begin{align}
\mathcal{H}= \begin{pmatrix}
H^+ & 0 \\
0 & H^-
\end{pmatrix}
\end{align}

where $H^\kappa=H^\kappa_{L} + H^\kappa_{R} + H_{T} + H_{B} + H_{N} +H_{C}$ where $\kappa=\pm 1$ labels the sectorial degree of freedom and $
    H_\nu^\kappa=-2t_0(\cos{k_x} +\cos{k_y}-\mu)\tau_z\sigma_0 +t_j(\sin{k_x}\cos{\alpha} +\sin{k_{y}}\sin{\alpha})\tau_0\sigma_z+\kappa J_{sd}\tau_z\sigma_x  -  \Delta_s(\cos{\phi_\nu}\tau_y + \sin{\phi_\nu}\tau_x)\sigma_y$.
Next, we express each component of the system in real space tight-binding representation. The NM region is modeled as a two-dimensional square lattice with lattice constant $a$, comprising of $N_x$ sites along the $x$-axis and $N_y$ sites along the $y$-axis. As a result, the widths of the left and right SC leads are $N_y a$, while the top and bottom SC leads have width $N_x a$. For convenience, the lattice sites in the NM region are labeled such that the coordinate $(0,0)$ denotes the bottom-left corner of the square barrier, as shown in Fig.[\ref{fig1aa}]. The resulting device geometry corresponds to a cross-shaped four terminal JJ mediated through a central NM barrier region. The individual components of the total Hamiltonian in Eq.~(\ref{eq1}) then read as follows
\begin{align}
\nonumber
    H_{\nu}^{\kappa} = &\sum_{\substack{\sigma, (i,j) \in \nu}} \big[ (4t_0-\mu) c^{\dagger}_{i,j,\sigma}c_{i,j,\sigma}+\sigma\Delta_{s}e^{i\phi_\nu} c^{\dagger}_{i,j,\sigma}c^{\dagger}_{i,j,\bar{\sigma}}\\& + \frac{t_j}{2i}\sigma~\big(\cos{\alpha}~c^{\dagger}_{i+1,j,\sigma}c_{i,j,\sigma} + \sin{\alpha}~c^{\dagger}_{i,j+1,\sigma}c_{i,j,\sigma}\big)\nonumber\\
    \nonumber
    & -t_0(c^{\dagger}_{i+1,j,\sigma}c_{i,j,\sigma}
    +c^{\dagger}_{i,j+1,\sigma}c_{i,j,\sigma})\\
    &+\kappa J_{sd}c^{\dagger}_{i,j,\sigma}c_{i,j,\bar{\sigma}}+\text{H.c.}\big] ,
\label{eq2}
\end{align}

\begin{align}
\nonumber
    H_{\mu} = &\sum_{\substack{\sigma, (i,j) \in \mu}} \big[(4t_0-\mu) c^{\dagger}_{i,j,\sigma}c_{i,j,\sigma} \nonumber\\
    &+\sigma\Delta_{p}e^{i\phi_\mu}( c^{\dagger}_{i,j+1,\sigma}c^{\dagger}_{i,j,\sigma} -c^{\dagger}_{i,j,\sigma}c^{\dagger}_{i,j+1,\sigma})\\
    &-t_0 (c^{\dagger}_{i+1,j,\sigma}c_{i,j,\sigma}+c^{\dagger}_{i,j+1,\sigma}c_{i,j,\sigma}) +\text{H.c.} \big],
\label{eq3}
\end{align}

\begin{align}
\nonumber
    H_N= &\sum_{\sigma,i ,j} \big[(4t_0-\mu + V_{G}) c^{\dagger}_{i,j,\sigma}c_{i,j,\sigma}\\
    &-t_0(c^{\dagger}_{i+1,j,\sigma}c_{i,j,\sigma} 
    +c^{\dagger}_{i,j+1,\sigma}c_{i,j,\sigma}) +\text{H.c.}\big],
    \label{eq4}
\end{align}
\begin{align}
\nonumber
    H_C&=\Big[t_l \sum_{\sigma,\,0\leq j< N_y}(c^{\dagger}_{-1,j,\sigma}c_{0,j,\sigma} + c^{\dagger}_{N_x+1,j,\sigma}c_{N_x,j,\sigma} ) \\
    &+ t_t\sum_{\sigma,\, 0\leq i < N_x}(c^{\dagger}_{i,-1,\sigma}c_{i,0,\sigma} + c^{\dagger}_{i,N_y+1,\sigma}c_{i,N_y,\sigma} )\Big] + \text{H.c.}
    \label{eq5}
\end{align}
Here, the notation $(i,j)\in\nu$ represents the lattice sites that belong to the left ($L$) or right ($R$) lead, corresponding to $i<0$ or $i \geq N_x$, with $0\leq j<N_y$. Similarly, $(i,j)\in\mu$ identifies lattice sites in the bottom ($B$) or top ($T$) leads, defined by $0 \leq i<N_x$ and $j<0$ or $j\geq N_y$, respectively. In the above Hamiltonian components, $c^{\dagger}_{i,j,\sigma}$ ($c_{i,j,\sigma}$) represents the creation (annihilation) operator of an electron with spin $\sigma= 1 (\bar \sigma=-1) $ at the lattice site $(i,j)$. The sectoral degree of freedom is labeled by the superscript $\kappa = \pm 1$ in Eq.~(\ref{eq2}), and enters only through the sign of the $sd$-exchange coupling  $J_{sd}$. The hopping amplitudes $t_l$ ($t_t$) quantify the transparency of the left and right (top and bottom) interfaces, and thus control the coupling between the superconducting leads and the NM region. For completeness, the matrix representation of the Hamiltonians  in Eqs.~(\ref{eq2})--(\ref{eq5}) in the Nambu basis is provided in Appendix~[\ref{AppA}].

\section{Algorithm}
\label{sec2}
In this section, we present the formalism used to evaluate the charge and spin currents flowing from each lead through the four-terminal junction. As discussed in Sec.~\ref{sec1}, the BdG Hamiltonian is block diagonal in the sectoral basis, allowing the problem to decompose into two independent sectors, $\kappa=\pm 1$. Consequently, all Green's functions are computed separately within each sector, and the physical current in each $\beta$ lead is obtained by summing the corresponding sectoral contributions, $\mathbf{I}_\beta^{c/s}=\sum_{\kappa=\pm1}\mathcal{I}_{\beta}^{c/s,\kappa}$.
The charge current is calculated from the time derivative of the number operator associated with each lead, evaluated across the microscopic lattice bonds at the junction-lead interfaces. The charge current flowing from lead $\beta$ is expressed as [\onlinecite{PhysRevResearch.2.023197,PhysRevB.110.014518,sharma2026pwavemagnetdrivenfieldfree,yqsg-xdg8,PhysRevB.104.134514,Sun_2009,PhysRevB.93.195302}]
\begin{align}
    \mathcal{I}_{\beta}^{c,\kappa}= -\frac{e}{h}\int dE\;\mathrm{Tr}\Big(\Gamma_{z}\hat{T}_{\beta}
    G^{<}_{N\beta}(E)+\mathrm{H.c.}\Big),
    \label{eq6}
\end{align}
where $\beta\in{L,R,T,B}$ labels the four leads, and $G^{<}_{N\beta}(E)$ denotes the nonlocal lesser Green's function evaluated at the corresponding junction--lead interface in the Nambu basis. The trace is taken over the Nambu, spin  degrees of freedom, and transverse lattice-sites. The operator $\Gamma_{z}=\tau_{z}\otimes\sigma_{0}$ accounts for the particle-hole structure by assigning opposite signs to the particle and hole components. The coupling
matrices at the interfaces are
$\hat{T}_{L/R(T/B)}=t_{l(t)}\,\mathbbm{1}_{N_y(N_x)}\otimes\tau_{z}\otimes
\sigma_{0}$, where $\mathbbm{1}_d$ is the identity matrix of dimension $d$.
Throughout this work, we adopt the convention that a positive (negative) current corresponds to a current flowing out of (into) the lead. The nonlocal lesser Green's function at the junction-lead interface is obtained from the fluctuation-dissipation theorem,
\begin{align}
G^{<}_{N\beta}(E) = -f(E)\big(G^{r}_{N\beta}(E) - G^{a}_{N\beta}(E)\big),
\label{eq7}
\end{align}
where $f(E)$ is the Fermi-Dirac distribution function. The nonlocal retarded (advanced) Green's function, $G^{r(a)}_{N\beta}(E)$, is given by
\begin{align}
    G^{r(a)}_{N\beta}(E)=G_{NN}^{r(a)}(E)\hat{T}_{\beta}^{\dagger}g_{\beta}^{r(a)}(E).
    \label{eq8}
\end{align}
Here, $G^{r(a)}_{NN}(E)$ denotes the boundary block of the retarded (advanced) Green's function of the normal region, evaluated at the lattice sites adjacent to the interface with lead $\beta$, and $g^{r(a)}_{\beta}(E)$ is the surface Green's function of the corresponding isolated lead. The latter is computed using the M\"{o}bius transformation [\onlinecite{PhysRevB.55.5266}]. Details of the calculation of $g^{r(a)}_{\beta}(E)$ and $G_{NN}^{r(a)}(E)$ are provided in Appendix~\ref{AppB}. The expression for the spin current follows directly from Eq.~(\ref{eq6}) by replacing the charge operator with the spin operator. Accordingly, the $z$-polarized spin current flowing from lead $\beta$ is
\begin{align}
    \mathcal{I}_{\beta}^{s,\kappa}=-\frac{1}{4\pi}\int dE\;\mathrm{Tr}\Big(\tilde\Gamma_{z}
    \hat{T}_{\beta}G^{<}_{N\beta}(E)+\mathrm{H.c.}\Big),
    \label{eq9}
\end{align}
where $\tilde{\Gamma}{z}=\tau_{z}\otimes\sigma_{z}$. We note that compared with the charge-current expression, the prefactor $e$ is replaced by the spin angular momentum factor $\hbar/2$.

\section{Symmetry Analysis}
\label{sec3}
In this section, we analyze the symmetry-imposed constraints on the charge and spin currents as functions of the superconducting phases and system parameters. We first define the transformation properties of the electron annihilation operator under the relevant symmetry operations as follows [\onlinecite{PhysRevB.104.134514}]

\begin{align}
    \nonumber
    \mathcal{T}c_{i,j,\sigma}\mathcal{T}^{-1}=& \sigma c_{i,j,\bar{\sigma}}
    \\
    \nonumber
    M_{xz}c_{i,j,\sigma}M_{xz}^{-1}=& \bar{\sigma} c_{i,-j,\bar{\sigma}}
    \\
    \nonumber
    M_{yz}c_{i,j,\sigma}M_{yz}^{-1}=& -i c_{-i,j,\bar{\sigma}}
    \\
    R_{x}c_{i,j,\sigma}R_x^{-1}=& ic_{i,j,\bar{\sigma}}
    \nonumber
    \\
    \nonumber
    R_yc_{i,j,\sigma}R_y^{-1}=& \sigma c_{i,j,\bar{\sigma}}
    \\
    R_zc_{i,j,\sigma}R_z^{-1}=& i\sigma c_{i,j,\sigma}.
    \label{eq10}
\end{align}
Here, $\mathcal{T}$ denotes the time-reversal operator, $M_{xz}$ and $M_{yz}$ are mirror reflections about the $xz$ and $yz$ planes, respectively, and $R_{x}$, $R_{y}$, and $R_{z}$ represent spin rotations by $180^{\circ}$ about the $x$, $y$, and $z$ axes. The corresponding matrix representations in the Nambu basis, together with the transformation properties of the individual terms appearing in Eqs.~(\ref{eq2}-\ref{eq4}), are summarized in Appendix~\ref{AppC}. In particular, some of these symmetry operations act only on the spin and spatial degrees of freedom and therefore leave the sector index $\kappa$ unchanged. 
However, certain symmetry operations reverse the sign of the exchange field, thereby interchanging the two sectors $H^{+}\leftrightarrow H^{-}$. We first examine the transformation of the total Hamiltonian under the spin rotation $R_{z}$ as,

\begin{align}
\nonumber
    R_{z}H^{\kappa}(&\frac{\phi_{l}}{2},-\frac{\phi_{l}}{2},\frac{\phi_{t}}{2},-\frac{\phi_{t}}{2},\alpha)R_{z}^{-1} \\
    \nonumber
    & =H^{\bar{\kappa}}(\frac{\phi_{l}}{2},-\frac{\phi_{l}}{2},\frac{\phi_{t}}{2} \pm\pi,-\frac{\phi_{t}}{2}\pm \pi, \alpha),\\
    &=H^{\bar{\kappa}}( \frac{(\phi_{l}\pm 2\pi)}{2},-\frac{(\phi_{l}\pm 2\pi)}{2},\frac{\phi_{t}}{2} ,-\frac{\phi_{t}}{2}, \alpha),
    \label{eq16}
\end{align}
where, the symbol $\pm$ indicates that the relation holds independently for either choice of sign. The second equality in Eq.~(\ref{eq16}) follows from the fact that the two sets of superconducting phases differ only by a global phase shift of $\pm\pi$, which corresponds to a gauge transformation. Following from the transformation of $H^{\kappa}$, the sectorial charge and spin current transform under $R_{z}$ as follows
\begin{align}
    \mathcal{I}_{\beta}^{s/c,\kappa}(\phi_{l},\phi_{t},\alpha)
    = \mathcal{I}_{\beta}^{s/c,\bar{\kappa}}(\phi_{l}\pm2\pi,\phi_{t},\alpha),
    \label{eq12}
\end{align}
for all values of $\phi_{l}$, $\phi_{t}$, and $\alpha$. Each sector is, therefore individually
$4\pi$ periodic in $\phi_{l}$. Equation~(\ref{eq12}) shows that the current in one sector is completely determined by that in the other through a $2\pi$ shift of the longitudinal phase. Consequently, each sector exhibits a $4\pi$-periodic current phase relation(CPR ) in $\phi_{l}$, while the total current, obtained by summing over the two sectors, is $2\pi$ periodic.
\begin{align}
\nonumber
    \mathbf{I}_{\beta}^{s/c}(2\pi+\phi_{l}) &= \mathcal{I}_{\beta}^{s/c,+} (2\pi+\phi_{l}) + \mathcal{I}_{\beta}^{s/c,-} (2\pi+\phi_{l})\\
    \nonumber
    &=\mathcal{I}_{\beta}^{s/c,-} (\phi_{l}) + \mathcal{I}_{\beta}^{s/c,+} (\phi_{l})\\
    &=\mathbf{I}_{\beta}^{s/c}(\phi_{l}),
\end{align}
where the second equality follows from Eq.~(\ref{eq12}). Next, we consider a combined symmetry operation $\mathcal{W}=M_{yz}M_{xz}\mathcal{T}R_{x}$, under which the Hamiltonian remains invariant, while the charge and spin currents transform as   
\begin{align}
\nonumber
    \mathcal{I}_{L(T)/R(B)}^{c,\kappa}\xrightarrow{M_{yz}} \mathcal{I}_{R(T)/L(B)}^{c, \kappa} \xrightarrow{M_{xz}} \mathcal{I}_{R(B)/L(T)}^{c,\bar{\kappa}} \\
    \xrightarrow{\mathcal{T}}- \mathcal{I}_{R(B)/L(T)}^{c,\kappa}\xrightarrow{R_{x}}- \mathcal{I}_{R(B)/L(T)}^{c,\kappa}\label{eq19}\\ \nonumber
    \mathcal{I}_{L(T)/R(B)}^{s,\kappa}\xrightarrow{M_{yz}} -\mathcal{I}_{R(T)/L(B)}^{s,\kappa} \xrightarrow{M_{xz}} \mathcal{I}_{R(B)/L(T)}^{s,\bar{\kappa}} \\
    \xrightarrow{\mathcal{T}} \mathcal{I}_{R(B)/L(T)}^{s,\kappa}\xrightarrow{R_{x}}- \mathcal{I}_{R(B)/L(T)}^{s,\kappa}
    \label{eq20}
\end{align}
Following the adopted sign convention, these relations indicate that the charge and spin currents leaving the left (top) lead are exactly balanced by those entering the right (bottom) lead, and vice versa. Consequently, no net current can flow between the horizontal and vertical transport channels. It is therefore sufficient to focus on the currents associated with the left and bottom leads, which we denote as the longitudinal ($\mathbf{I}_{l}^{c/s}$) and transverse ($\mathbf{I}_{t}^{c/s}$) currents, respectively. In the following section, we introduce asymmetric interface couplings and a random on-site potential. Both perturbations break the $M_{yz}$ and $M_{xz}$ mirror symmetries, thereby removing the symmetry constraints discussed above. Under these conditions, the longitudinal and transverse currents are defined as follows:
\begin{align}
    \mathbf{I}_{l}^{c/s}=\frac{\mathbf{I}_{L}^{c/s}-\mathbf{I}_{R}^{c/s}}{2},\qquad
    \mathbf{I}_{t}^{c/s}=\frac{\mathbf{I}_{B}^{c/s}-\mathbf{I}_{T}^{c/s}}{2},
    \label{eq21}
\end{align}
which reduce to $\mathbf{I}_{L}^{c/s}$ and $\mathbf{I}_{B}^{c/s}$ when the symmetry $\mathcal{W}=M_{yz}M_{xz}\mathcal{T}R_{x}$ is present. We next consider the combined symmetry operation $\mathcal{T}R_{z}$, under which the Hamiltonian transforms as
\begin{align}
\nonumber
    \mathcal{T}R_{z}&H^{\kappa}\Big(\frac{\phi_{l}}{2},-\frac{\phi_{l}}{2},
    \frac{\phi_{t}}{2},-\frac{\phi_{t}}{2},\alpha\Big)R_{z}^{-1}\mathcal{T}^{-1}\\
    &=H^{\kappa}\Big(-\frac{\phi_{l}}{2},\frac{\phi_{l}}{2},
    -\frac{\phi_{t}}{2}\pm\pi,\frac{\phi_{t}}{2}\pm\pi,\alpha\Big).
    \label{eq17}
\end{align}
Two special cases arise when the transverse phase is fixed at $\phi_{t}=0$ and $\phi_{t}=\pi$. For $\phi_{t}=0$, the transformed phase configuration differs from the original parametrization only by a global gauge transformation, yielding
$H^{\kappa}\big(\frac{2\pi-\phi_{l}}{2},-\frac{2\pi-\phi_{l}}{2},0,0,\alpha\big)$ whereas for $\phi_{t}=\pi$, the Hamiltonian transforms to $H^{\kappa}\big(-\frac{\phi_{l}}{2},\frac{\phi_{l}}{2},\frac{\pi}{2},
-\frac{\pi}{2},\alpha\big)$. Since the charge (spin) current is odd (even) under the combined symmetry operation $\mathcal{T}R_{z}$, the corresponding current-phase relations satisfy
\begin{align}
\nonumber
    \mathcal{I}_{l/t}^{c(s),\kappa}(\phi_{l},\phi_{t}=0)
    &=-(+)\,\mathcal{I}_{l/t}^{c(s),\kappa}(2\pi-\phi_{l},\phi_{t}=0),\\
    \mathcal{I}_{l/t}^{c(s),\kappa}(\phi_{l},\phi_{t}=\pi)
    &=-(+)\,\mathcal{I}_{l/t}^{c(s),\kappa}(-\phi_{l},\phi_{t}=\pi).
    \label{eq18}
\end{align}
Eq.~(\ref{eq18}) implies that the maximum and minimum values of both the charge and spin currents are of equal magnitude. For the charge (spin) current, the negative (positive) sign reflects the odd (even) parity of the CPR at $\phi_t=0$ and $\phi_t=\pi$. Consequently, the critical currents in the forward and reverse directions are identical, prohibiting a finite CDE at these transverse phases. The relation given in Eq. (\ref{eq18}) is also followed for $\phi_{t}=-\pi$ which can be obtained using opposite sign of phase transformation in Eq.~(\ref{eq17}). In the next section, we present the numerical results which fully confirm this symmetry-imposed constraint. A finite CDE emerges only when $\phi_t$ deviates from $0$ and $\pi$, where the odd-parity relation is no longer protected by symmetry.

Additional symmetry constraints arise when the transverse phase and the crystallographic orientation are fixed to specific values. We first consider the combined symmetry operation $\mathcal{X}=M_{yz}\mathcal{T}R_{x}$ at $\phi_{t}=0$ and $\alpha=\pi/2$. Under this operation, the Hamiltonian remains invariant while the two sectoral Hamiltonians are interchanged as follows
\begin{align}
\nonumber
    \mathcal{X}H^{\kappa}\Big(\frac{\phi_{l}}{2},-\frac{\phi_{l}}{2},0,0,\alpha=\pi/2\Big)
    \mathcal{X}^{-1}
    =\\H^{\bar{\kappa}}\Big(\frac{\phi_{l}}{2},-\frac{\phi_{l}}{2},0,0,\alpha=\pi/2\Big),
    \label{eq28}
\end{align}
The corresponding charge and spin currents transform as
\begin{align}
    \mathcal{I}^{c,\kappa}_{l(t)}(\phi_{l})&=+(-)\,\mathcal{I}_{l(t)}^{c,\bar{\kappa}}(\phi_{l}),\nonumber\\
    \mathcal{I}^{s,\kappa}_{l(t)}(\phi_{l})&=-(+)\,\mathcal{I}_{l(t)}^{s,\bar{\kappa}}(\phi_{l}).\label{eq29}
\end{align}
Since the total current is obtained by adding the two sectors, Eq.~(\ref{eq29}) immediately implies that the longitudinal spin current and the transverse charge current vanish, i.e., $\mathbf{I}_{l}^{s}=\mathbf{I}_{t}^{c}=0$. Consequently, the transverse current at $\phi_{t}=0$ is a pure spin current. Next, for $\phi_{t}=0$ and $\alpha=n\pi$, where $n\in \{0,1,2,\dots\} $, the operator $M_{yz}\mathcal{T}$ leaves the Hamiltonian invariant similar to Eq.~(\ref{eq28}), and the currents satisfy
\begin{align}
    \mathcal{I}^{c/s,\kappa}_{l(t)}(\phi_{l})=+(-)\,\mathcal{I}_{l(t)}^{c/s,\bar{\kappa}}(\phi_{l}).
    \label{eq31}
\end{align}
Again, summing over the two sectors yields $\mathbf{I}_{t}^{c}=\mathbf{I}_{t}^{s}=0$ implying that neither charge nor spin current can flow in the transverse direction.

We now turn to the case of a transverse phase difference $\phi_{t}=\pi$ and consider the two crystallographic orientations of the PM, $\alpha=0$ and $\pi/2$. We first examine the combined symmetry operation $\mathcal{Y}=M_{yz}\mathcal{T}R_{y}$, which, for $\phi_{t}=\pi$ and $\alpha=\pi/2$, leaves the Hamiltonian invariant:
\begin{align}
\nonumber
    \mathcal{Y}H^{\kappa}\Big(\frac{\phi_{l}}{2},&-\frac{\phi_{l}}{2},
    \frac{\pi}{2},-\frac{\pi}{2},\alpha=\pi/2\Big)\mathcal{Y}^{-1}\\
\nonumber
    &=H^{\kappa}\Big(\frac{\phi_{l}}{2},-\frac{\phi_{l}}{2},
    -\frac{\pi}{2}+\pi,\frac{\pi}{2}-\pi, \alpha=\pi/2\Big)\\
    &=H^{\kappa}\Big(\frac{\phi_{l}}{2},-\frac{\phi_{l}}{2},
    \frac{\pi}{2},-\frac{\pi}{2},\alpha=\pi/2\Big),
    \label{eq24}
\end{align}
The invariance follows because the spin rotation $R_{y}$ reverses the sign of the equal-spin triplet order parameter, thereby shifting each superconducting phase $\phi_{l/t}$ by $\pi$, while both $M_{yz}$ and $\mathcal{T}$ reverse the longitudinal phase difference. The corresponding charge and spin currents transform as
\begin{align}
    \mathcal{I}^{c,\kappa}_{l(t)}(\phi_{l})&=+(-)\,\mathcal{I}_{l(t)}^{c,\kappa}(\phi_{l}),\label{eq25}\\
    \mathcal{I}^{s,\kappa}_{l(t)}(\phi_{l})&=-(+)\,\mathcal{I}_{l(t)}^{s,\kappa}(\phi_{l})\label{eq26},
\end{align}
These relations immediately imply that
$\mathcal{I}_{t}^{c,\kappa}=\mathcal{I}_{l}^{s,\kappa}=0$ for all
$\phi_{l}$. Since $\mathcal{Y}$ preserves the sector index, these constraints hold independently within each sector. Consequently, for $\phi_{t}=\pi$ and $\alpha=\pi/2$, the transverse current is a pure spin current.
An analogous symmetry constraint arises for the complementary crystallographic orientation. For $\alpha=n\pi$ and $\phi_{t}=\pi$, the combined symmetry operation $\mathcal{Z}=M_{yz}\mathcal{T}R_{z}$ likewise leaves the Hamiltonian invariant through the same sequence of phase transformations as in Eq.~(\ref{eq24}), while the currents satisfy
\begin{align}
    \mathcal{I}^{c/s,\kappa}_{l(t)}(\phi_{l})=+(-)\,\mathcal{I}_{l(t)}^{c/s,\kappa}(\phi_{l}).
    \label{eq27}
\end{align}
It follows that $\mathcal{I}_{t}^{c,\kappa}=\mathcal{I}_{t}^{s,\kappa}=0$, which implies that both the transverse charge and spin currents vanish identically within each sector. Finally, we emphasize that the numerical results presented in the following section are in complete agreement with these symmetry-imposed constraints.

\section{Results}
\label{sec4}
In this section, we present numerical results demonstrating that the interplay between superconducting phase differences and PM anisotropy gives rise to  a fully non-local transverse spin and charge diode response with $100\%$ efficiency. The parameters are chosen as $\mu=2t_{0}$, $t_{0}=1$, $t_{j}=0.5t_{0}$, $J_{sd}=0.2t_{0}$, and $\Delta_{p}=10\Delta_{s}=0.05t_{0}$.  The order of magnitude difference between $\Delta_{s}$ and $\Delta_{p}$ is motivated by the fact that the $s$-wave pairing gap in the PM is proximity induced rather than intrinsic. Therefore, its magnitude is reduced relative to that of the parent SC as a result of finite interface coupling. Nevertheless, later, we examine various values of gap amplitudes explicitly. The system size is fixed at $N_{x}\times N_{y}=5\times5$, placing the device in the short-junction regime, and we assume transparent interfaces with $t_{l}=t_{t}=t_{0}$, unless otherwise specified.

\begin{figure}[t]
\begin{tabular}{c c}
    \centering
    \includegraphics[width=0.472\linewidth]{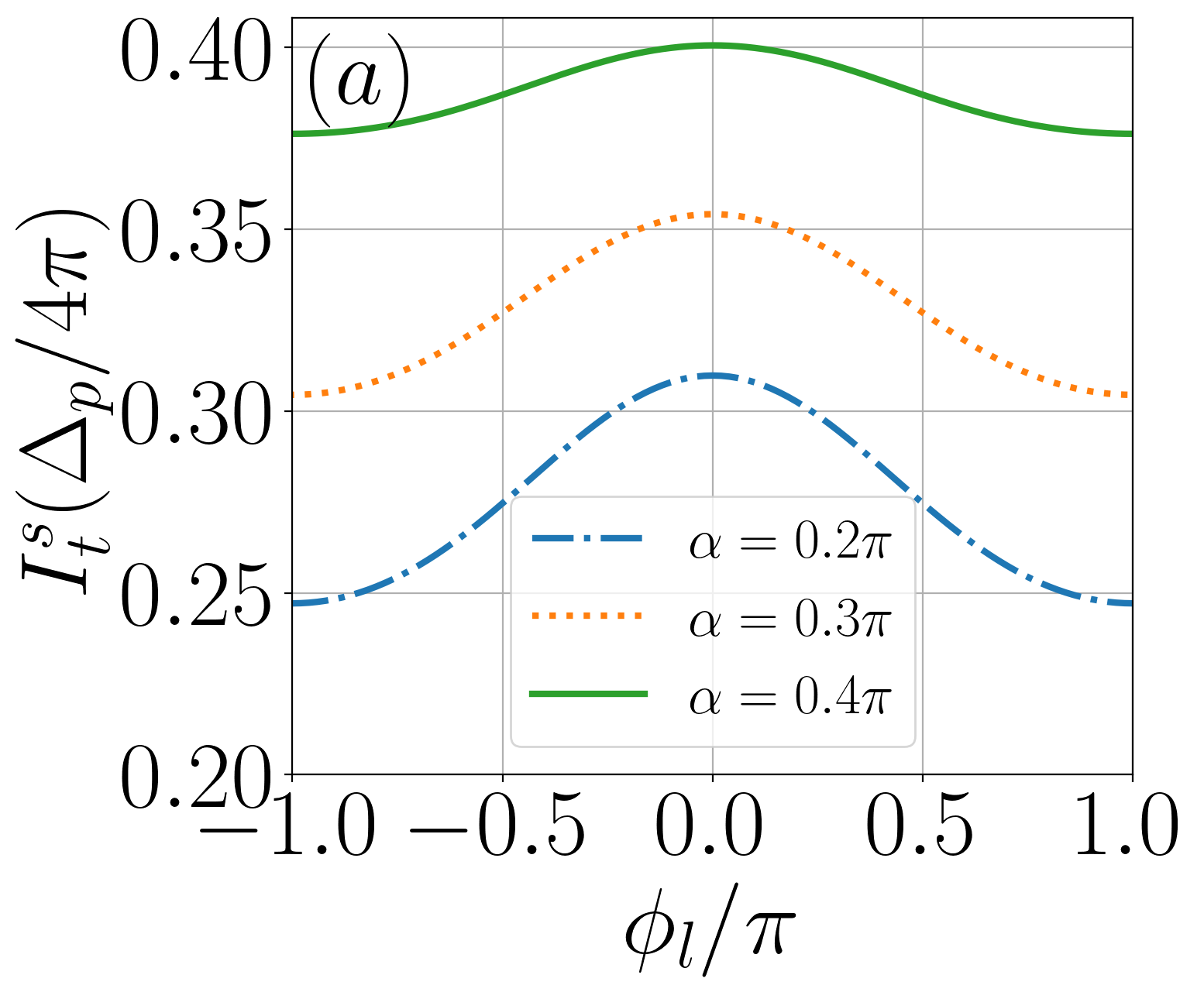}
    &\includegraphics[width=0.5\linewidth]{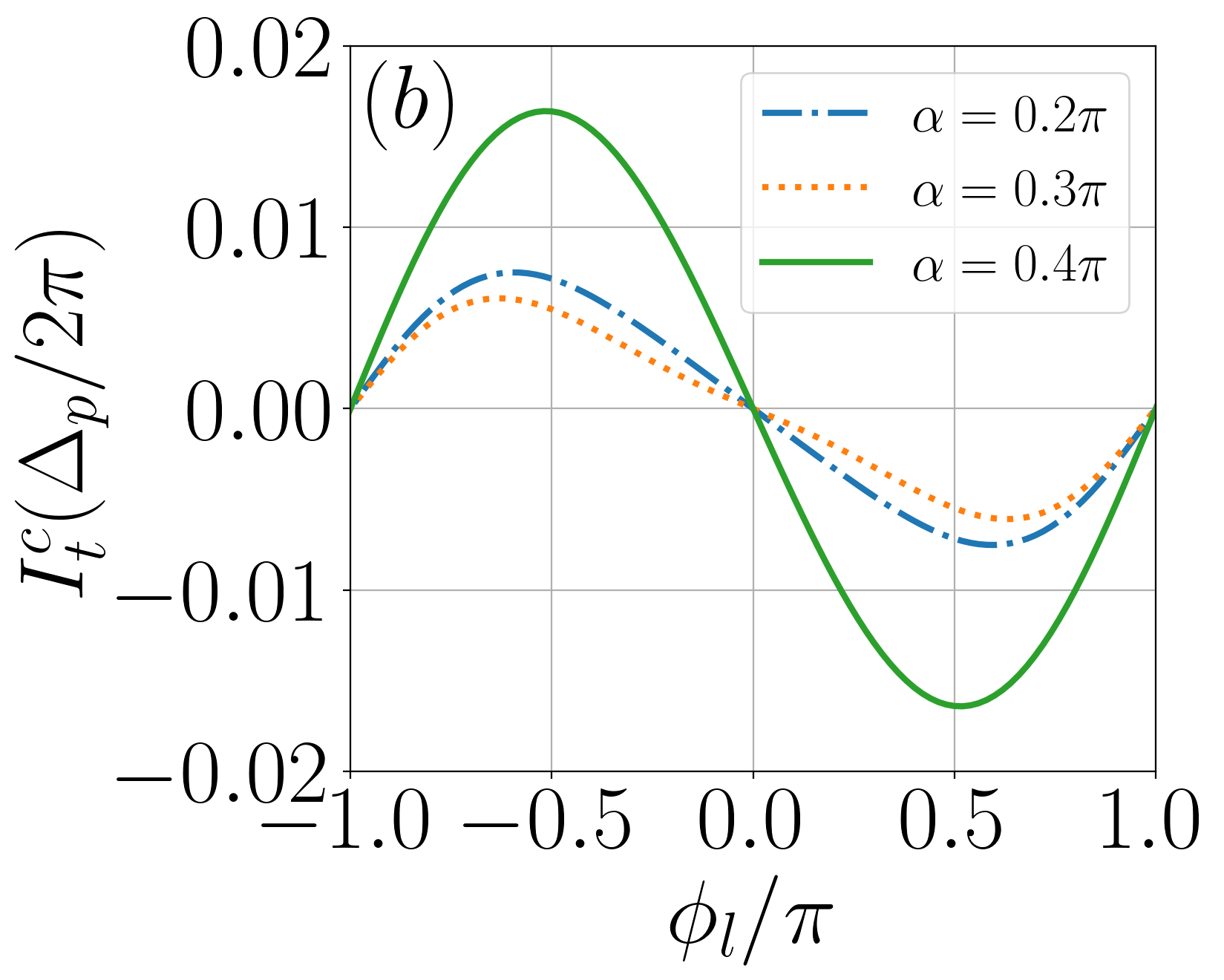}
    \end{tabular}
    \caption{Transverse (a)  spin current, $I^s_t$ (b) charge current, $I^{c}_t$ as functions of the longitudinal phase difference $\phi_{l}$ for crystallographic orientations $\alpha=0.2\pi$ (blue), $\alpha=0.3\pi$ (orange), and $\alpha=0.4\pi$ (green). The calculations are performed for $\phi_{t}=0$, $V_{G}=0$, $t_{l}=t_{t}=t_{0}$, $\mu=2t_{0}$, $t_{0}=1$  $t_{j}=0.5t_{0}$, $\Delta_{s}=0.005t_{0}$, $\Delta_{p}=0.05t_{0}$, $J_{sd}=0.2t_{0}$, and $N_{x}=N_{y}=5$.}
    \label{fig1}
\end{figure}

We begin by evaluating the CPRs of the transverse charge and spin currents as functions of the longitudinal phase difference $\phi_{l}$ using Eqs.~(\ref{eq6}) and (\ref{eq9}), respectively. To quantify the degree of nonreciprocity in the CPR, we define the diode efficiency of transverse spin (charge) as $\eta^{s(c)}=\frac{|{I^{s(c)}_{t}}^{+}| - |{I^{s(c)}_{t}}^{-}|}{|{I^{s(c)}_{t}}^{+}|+|{I^{s(c)}_{t}}^{-}|}$, where ${I^{s(c)}_{t}}^{+}$ and ${I^{s(c)}_t}^{-}$ denote the positive and negative critical transverse spin (charge) currents, respectively, obtained from the CPR. According to this definition, the absence of either the positive or negative critical current over the entire range of $\phi_l$ corresponds to a perfect $100\%$ diode effect. Figs.~\ref{fig1}(a) and \ref{fig1}(b) show the transverse spin and charge currents, respectively, for a fixed transverse phase difference $\phi_t=0$ and crystallographic orientations $\alpha=0.2\pi$, $0.3\pi$, and $0.4\pi$. The transverse spin current, $I_{t}^{s}$, retains the same sign over the entire range of $\phi_{l}$, demonstrating perfect nonreciprocity and hence a 100\% transverse SDE. In contrast, the transverse charge current is reciprocal, yielding zero charge diode efficiency. Furthermore, spin and charge CPRs exhibit even and odd parity with respect to $\phi_{l}$, respectively, in agreement with the symmetry constraints derived in Eq.~(\ref{eq18}). 

\begin{figure}[t]
\begin{tabular}{c c}
    \centering
    \includegraphics[width=0.5\linewidth]{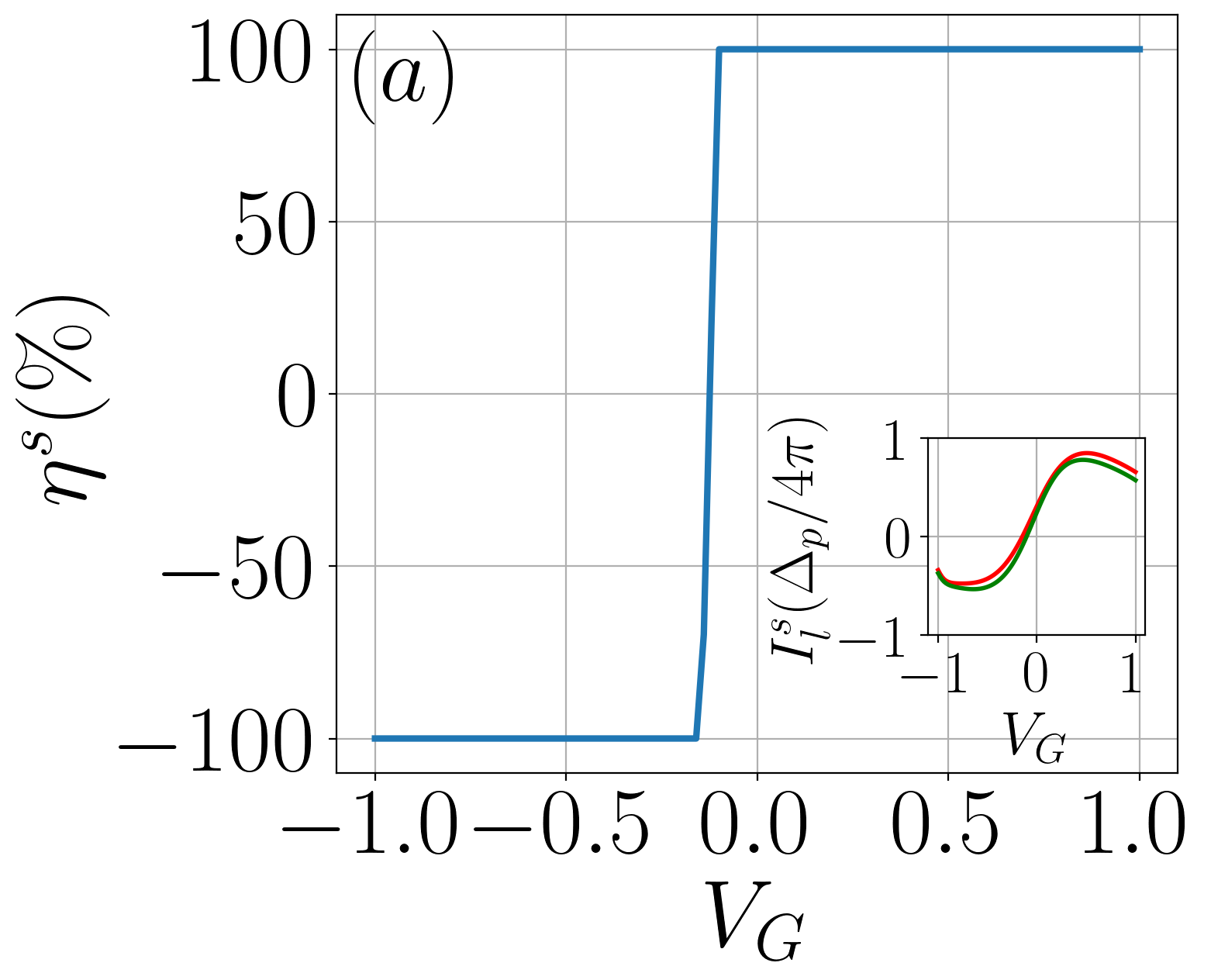}
    &\includegraphics[width=0.5\linewidth]{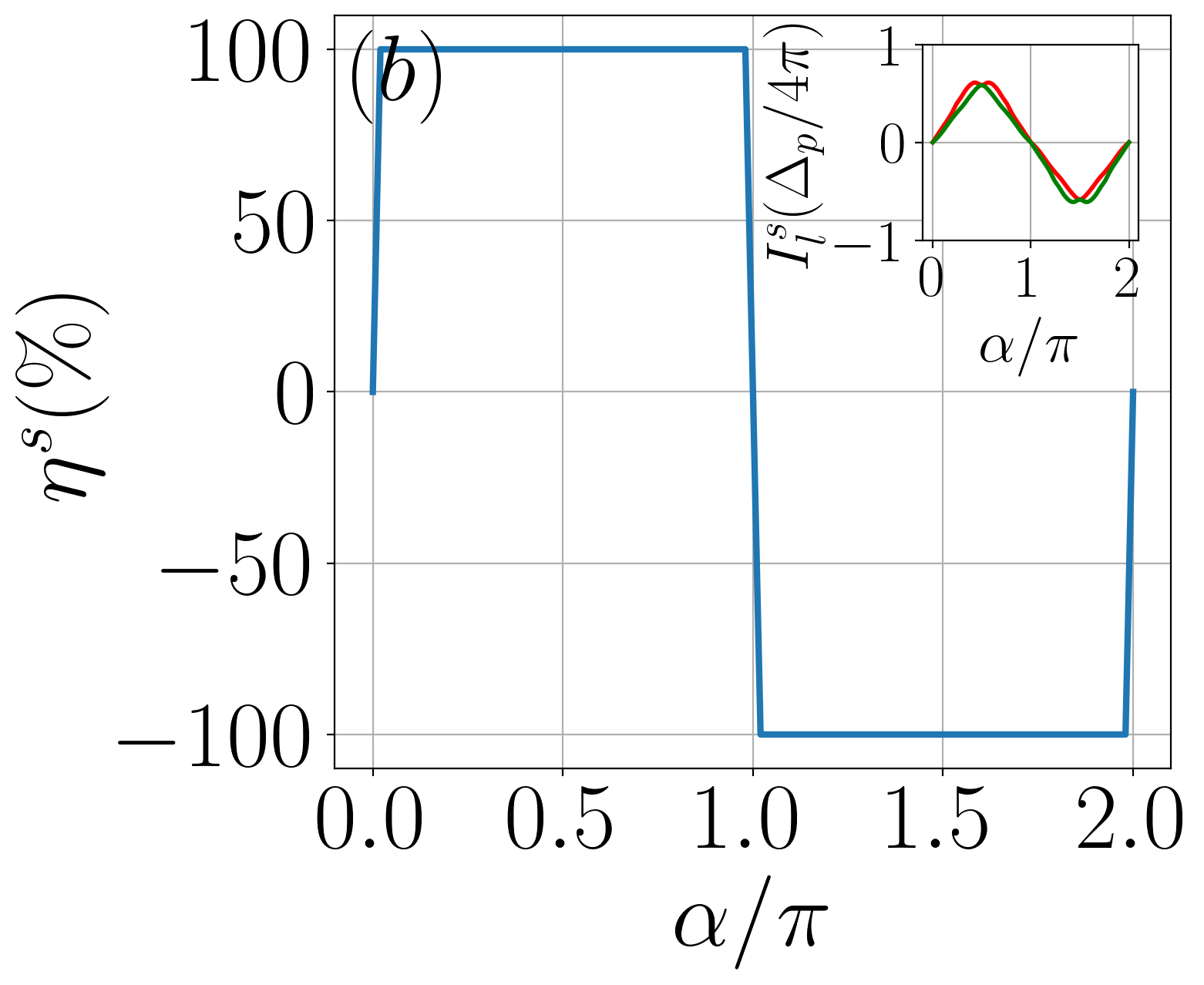}
    \end{tabular}
    \caption{ SDE $\eta^{s}$, as function of (a) gate potential, $V_{G}$ at $\phi_{t}=0$ and $\alpha=0.2\pi$ and (b) lobe angle of PM, $\alpha$ at $V_{G}=0$ and $\phi_{t}=0$ . The insets display maximum (red) and minimum (green) current as a function of $V_{G}$ and $\alpha$, respectively in the entire range of $\phi_{l}$. The identical sign of the maximum and minimum current values for every $V_G$ and $\alpha$ results in a perfect 100\% diode efficiency. Other system parameters are same as used in Fig [\ref{fig1}].}
    \label{fig2}
\end{figure}

Next, we investigate the tunability of the transverse SDE through the gate potential $V_{G}$, which effectively controls the chemical potential of the normal region. Fig.~\ref{fig2}(a) shows the SDE, $\eta^{s}$, as a function of $V_{G}$ for $\phi_{t}=0$ and $\alpha=0.2\pi$. Remarkably, over a wide range of gate voltages, the transverse spin current remains completely polarized in a single direction, corresponding to a perfect $100\%$ SDE. Also, a small variation in $V_{G}$ reverses the polarity of the diode efficiency, causing $\eta^{s}$ to switch abruptly from $+100\%\rightarrow-100\%$, resulting in an approximately step-like dependence on the gate potential. This shows that $V_{G}$ provides a simple and experimentally accessible means of electrically controlling the direction of nonlocal spin transport. Further, we show the dependence of the transverse SDE on the PM crystallographic orientation via Fig.~\ref{fig2}(b) which illustrates $\eta^{s}$ as a function of $\alpha$ for $V_{G}=0$ and $\phi_{t}=0$. The preferred direction of the SDE can be reversed by rotating the PM lobe angle, with the diode efficiency undergoing a sharp sign reversal at $\alpha = n\pi$, (where $\,n = 0, 1, 2, 3, \dots$). Although the spin current evolves continuously with $\alpha$, the diode efficiency changes abruptly at $\alpha = n\pi$ because both critical currents simultaneously vanish at these symmetry points. This behavior is a direct consequence of the symmetry constraint in Eq.~(\ref{eq31}), which requires both the transverse spin and charge currents to vanish identically for $\phi_{t}=0$ and $\alpha=n\pi$. Furthermore, at $\alpha=\pi/2$, Eq.~(\ref{eq29}) enforces the transverse charge current to vanish identically, implying that the observed perfect diode effect is purely of spin origin.

\begin{figure}[h]
    \centering

    \begin{subfigure}{0.237\columnwidth}
        \centering
        \includegraphics[width=\linewidth]{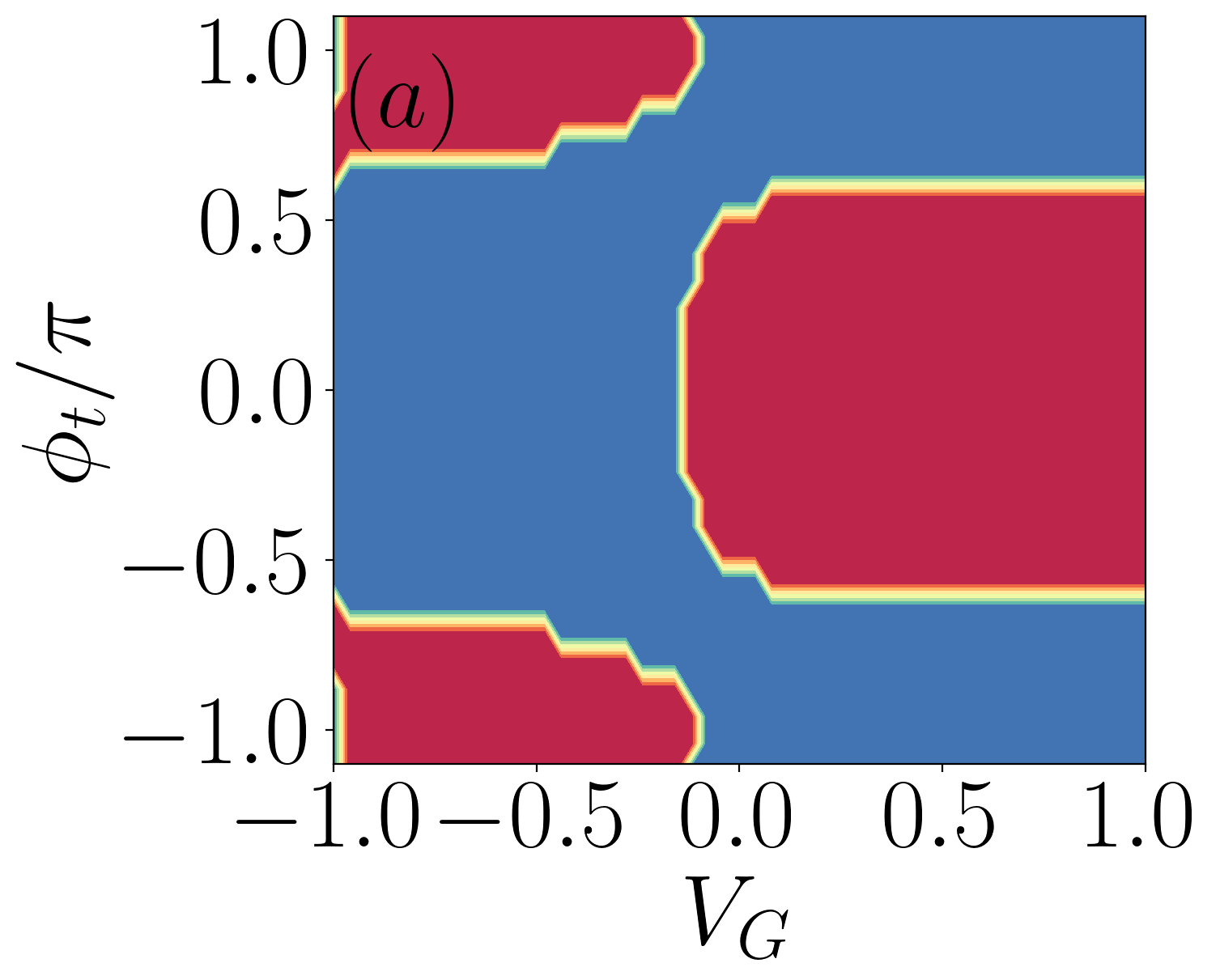}
    \end{subfigure}
    \hfill
    \begin{subfigure}{0.218\columnwidth}
        \centering
        \includegraphics[width=\linewidth]{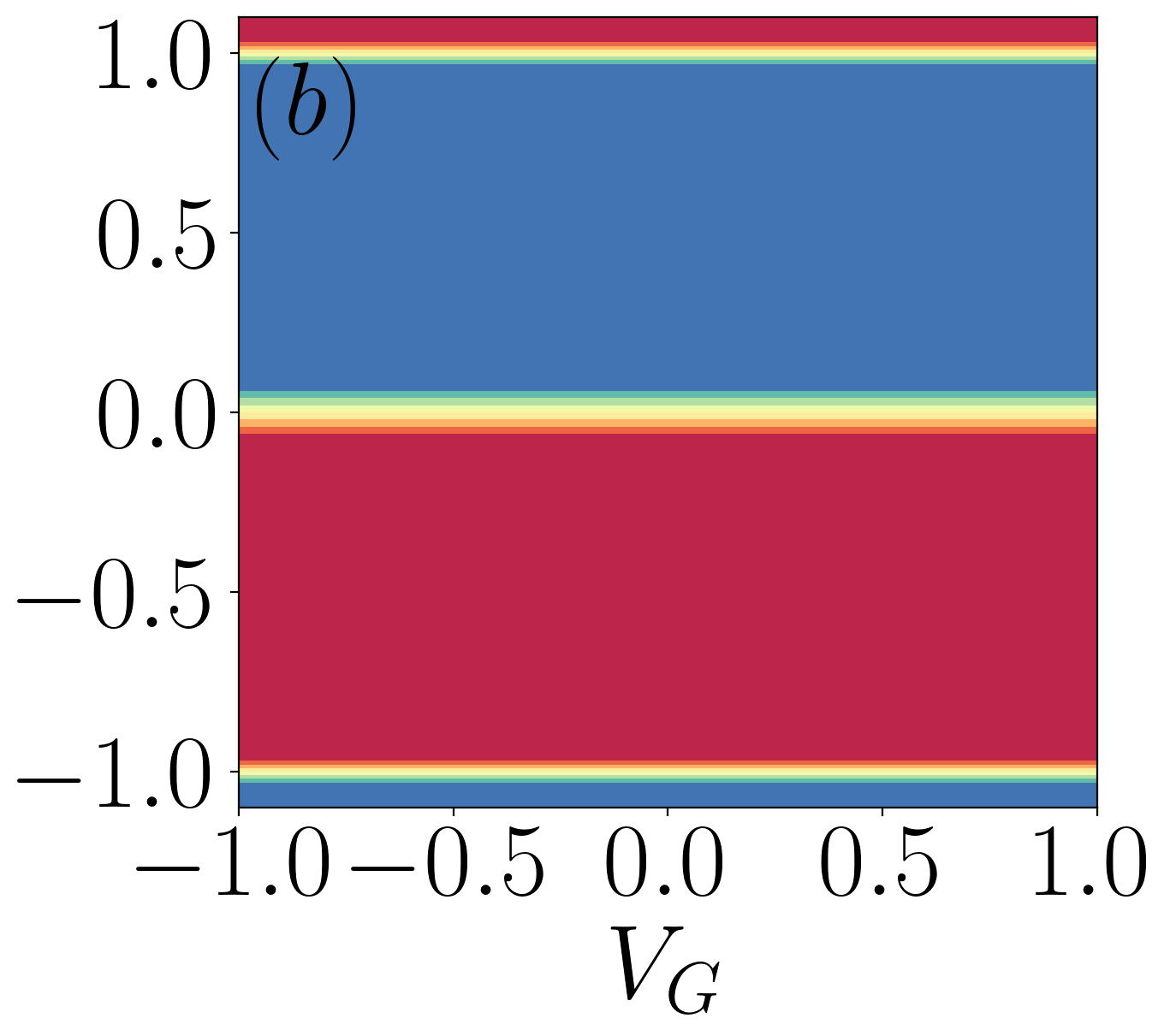}
    \end{subfigure}
    \hfill
    \begin{subfigure}{0.218\columnwidth}
        \centering
        \includegraphics[width=\linewidth]{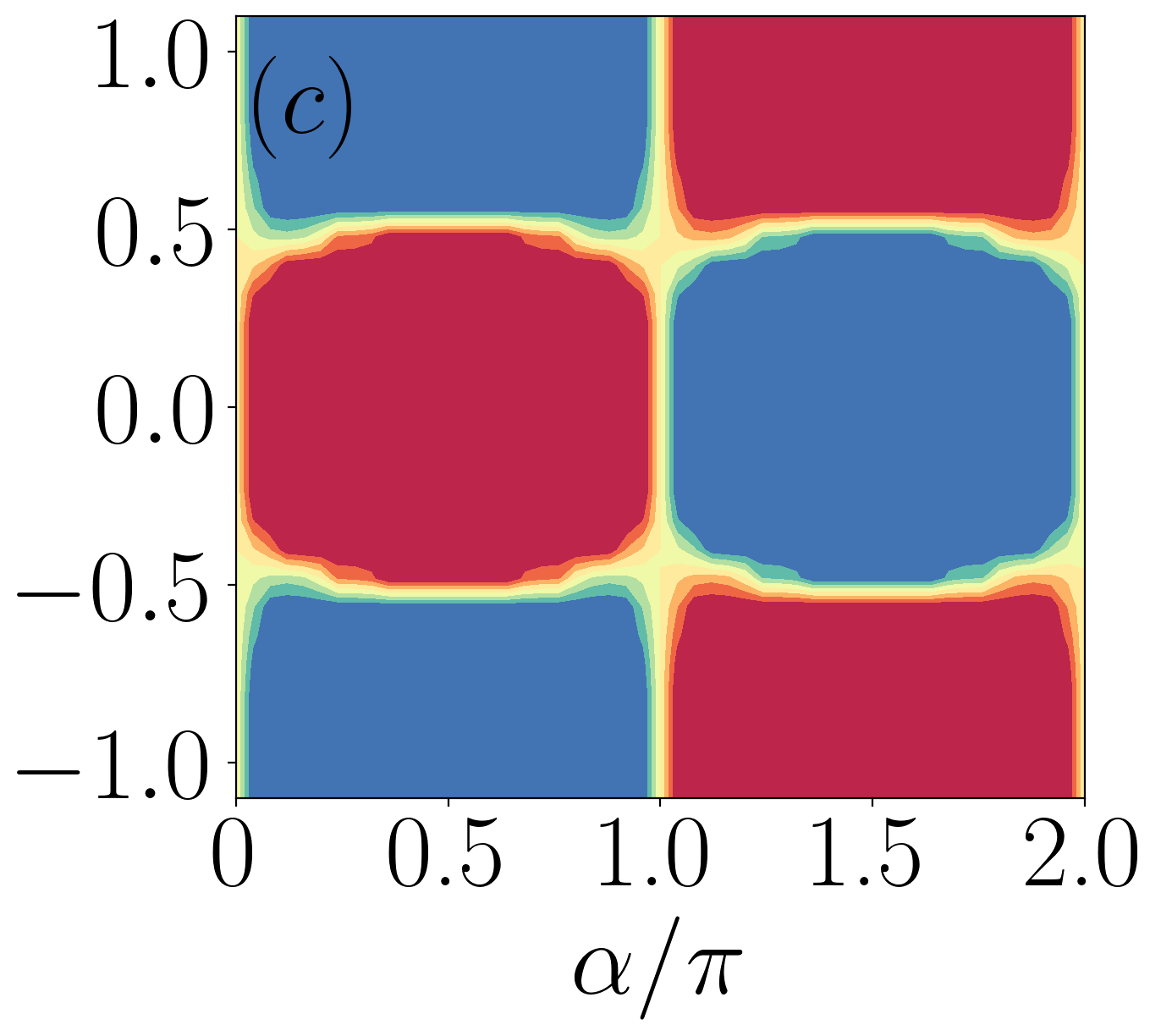}
    \end{subfigure}
        \hfill
    \begin{subfigure}{0.263\columnwidth}
        \centering
        \includegraphics[width=\linewidth]{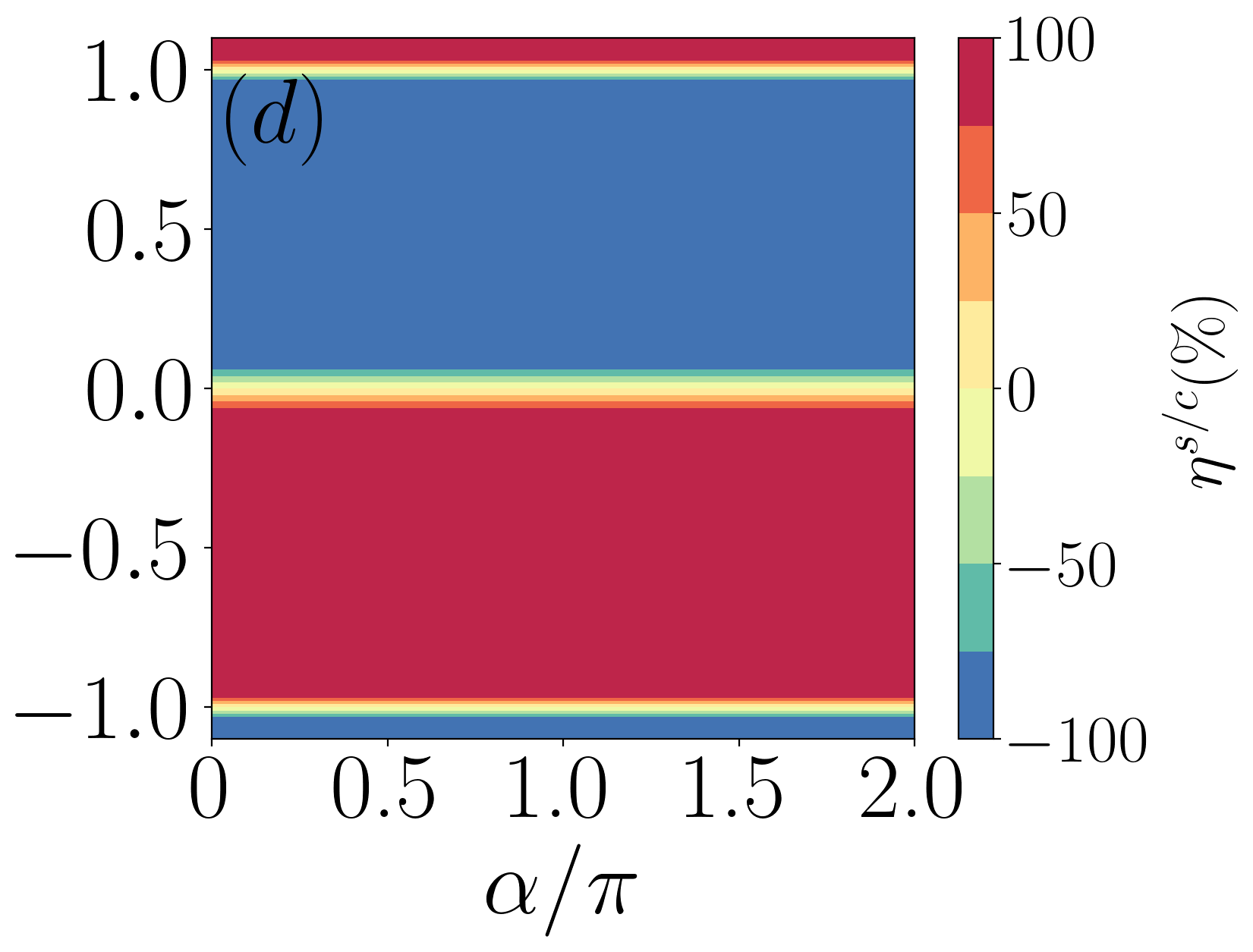}
    \end{subfigure}
    \caption{Contour plots of the SDE and CDE, $\eta^{s}$ [(a),(c)] and $\eta^{c}$ [(b),(d)], respectively, with the color scale representing these diode efficiencies. The horizontal axis corresponds to the gate potential $V_{G}$ in panels (a) and (b), and to the PM crystallographic angle $\alpha$ in panels (c) and (d), while the vertical axis denotes the transverse phase difference $\phi_{t}$ in all panels. All other parameters are the same as those used in Fig.~\ref{fig1}.}
    \label{fig4}
\end{figure}

To generalize our findings beyond the special case of $\phi_{t}=0$,  we present contour plots of the SDE ($\eta^{s}$) and  CDE ($\eta^{c}$) in  Figs.~\ref{fig4}(a) and \ref{fig4}(b), respectively, as functions of the gate potential $V_{G}$ and transverse phase difference $\phi_{t}$ for a representative lobe angle $\alpha=0.2\pi$. Fig.~\ref{fig4}(a) shows that SDE exhibits a rich dependence on both $V_{G}$ and $\phi_{t}$. For most of the parameter space, $\eta^{s}$ remains close to $\pm100\%$, while undergoing an abrupt, step-like reversal of polarity near $\phi_{t}\approx\pm0.6\pi$. Consequently, the direction of the transverse SDE can be controlled either electrically through the gate potential or phase coherently through the transverse superconducting phase difference. In contrast, the CDE is governed primarily by the transverse phase. As dictated by the symmetry relation in Eq.~(\ref{eq18}), the transverse charge current $I_t^{c}(\phi_l)$ is an odd function of $\phi_l$ at $\phi_t=0,\pm\pi$ and independent of $\alpha$. This gives rise to  $\eta^{c}=0$ along these symmetry-protected lines [see Fig.~\ref{fig4}(b)], although the charge current itself remains finite. Away from these lines, the transverse charge current becomes highly nonreciprocal, giving rise to a perfect CDE ($100\%$) over most of the $(V_{G},\phi_{t})$ parameter space. Unlike SDE, $\eta^{c}$ is only weakly affected by the gate potential and instead exhibits abrupt polarity reversals upon crossing each of the symmetry points $\phi_{t}=0,\pm\pi$. These results identify the transverse phase difference as the primary control parameter for the direction of the nonlocal CDE.

We next fix the gate voltage at $V_G=0$ and investigate the dependence of the diode efficiencies on the PM crystallographic orientation. Figs. \ref{fig4}(c) and  (d) show contour plots of the transverse
SDE and CDE in the $(\alpha$-$\phi_{t})$ plane, respectively. The transverse SDE exhibits an almost perfect efficiency over a wide parameter range, with the polarity reversing abruptly at $\alpha=n\pi$. This behavior originates from the fact that a longitudinal supercurrent flowing through the normal region is accompanied by a transverse spin-polarized current, as recently demonstrated in Ref.~[\onlinecite{salehi2025transversespinsupercurrentpwave}]. This arises from the spin-polarized anisotropic Fermi surface of the PM, and the resulting current reverses it's direction when lobe angle is rotated by $180^{\circ}$.
The transverse CDE, shown in Fig.~\ref{fig4}(d), likewise remains close to $100\%$  throughout most of the parameter space and is essentially independent of the crystallographic angle. It undergoes an abrupt polarity reversal upon crossing the symmetry-protected lines $\phi_{t}=0,\pm\pi$.
\section{ROBUSTNESS AND STABILITY OF DIODE EFFICIENCY}
\label{sec5}
\begin{figure}[b]
    \centering

    \begin{subfigure}{0.237\columnwidth}
        \centering
        \includegraphics[width=\linewidth]{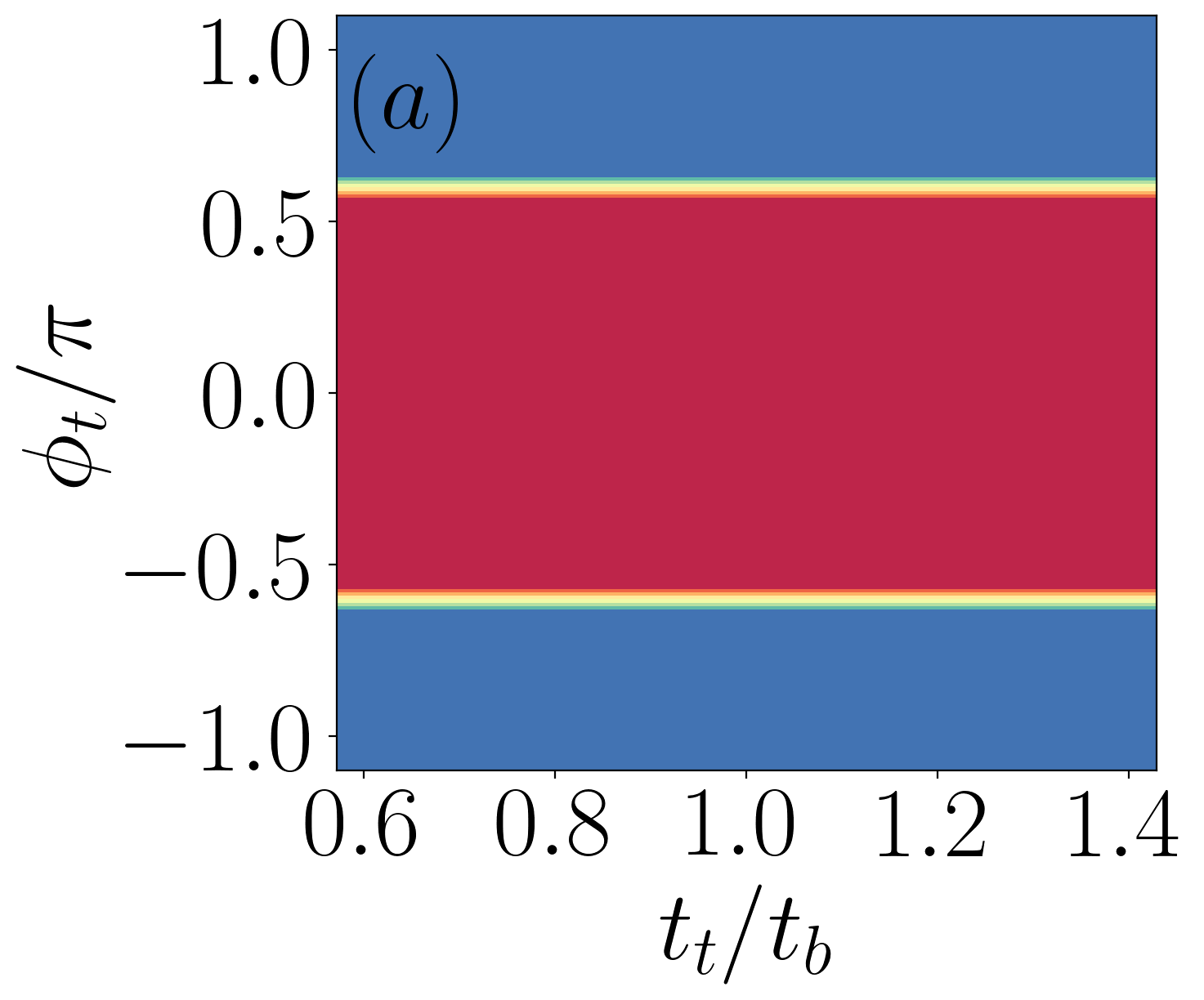}
    \end{subfigure}
    \hfill
    \begin{subfigure}{0.218\columnwidth}
        \centering
        \includegraphics[width=\linewidth]{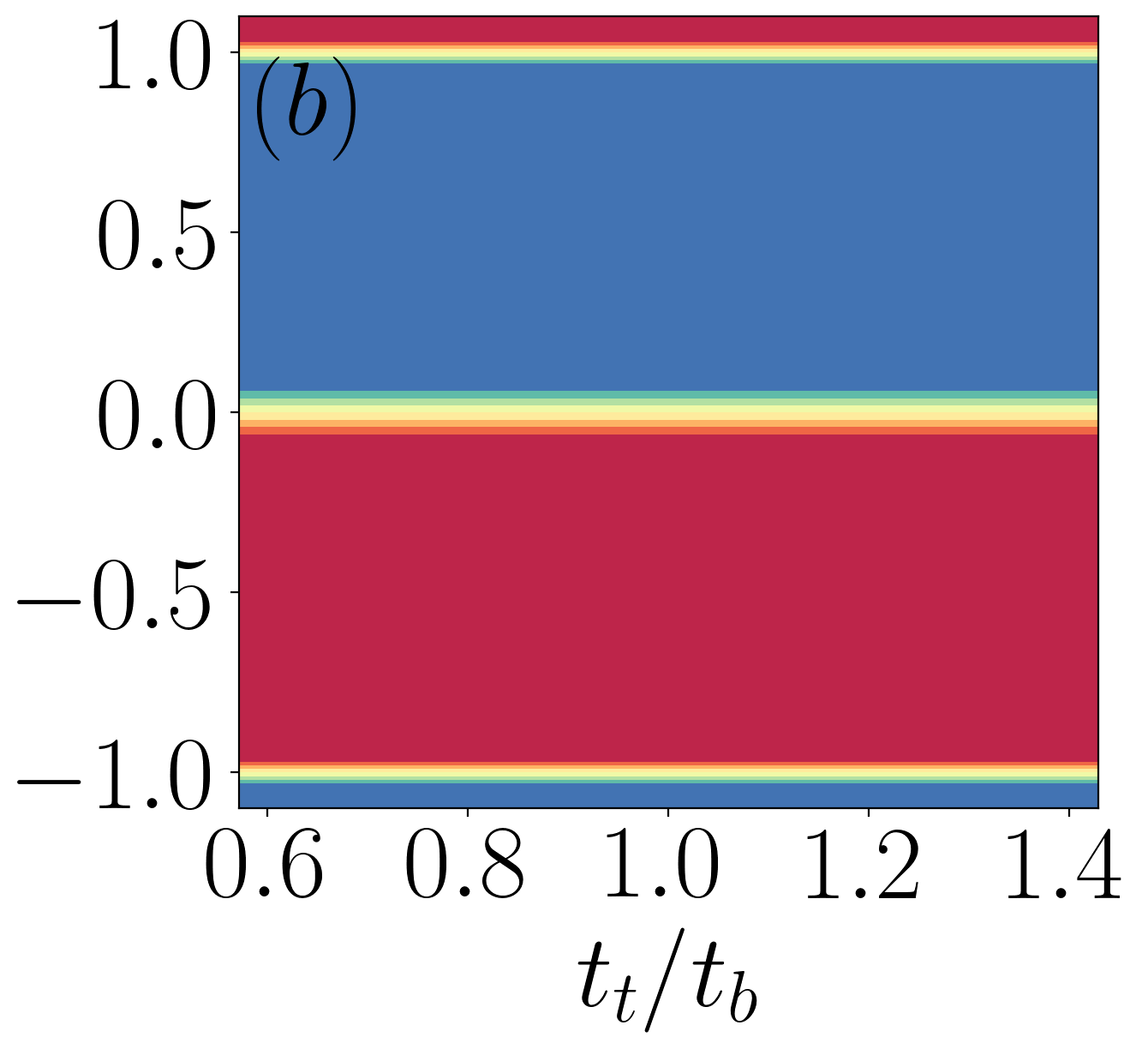}
    \end{subfigure}
    \hfill
    \begin{subfigure}{0.218\columnwidth}
        \centering
        \includegraphics[width=\linewidth]{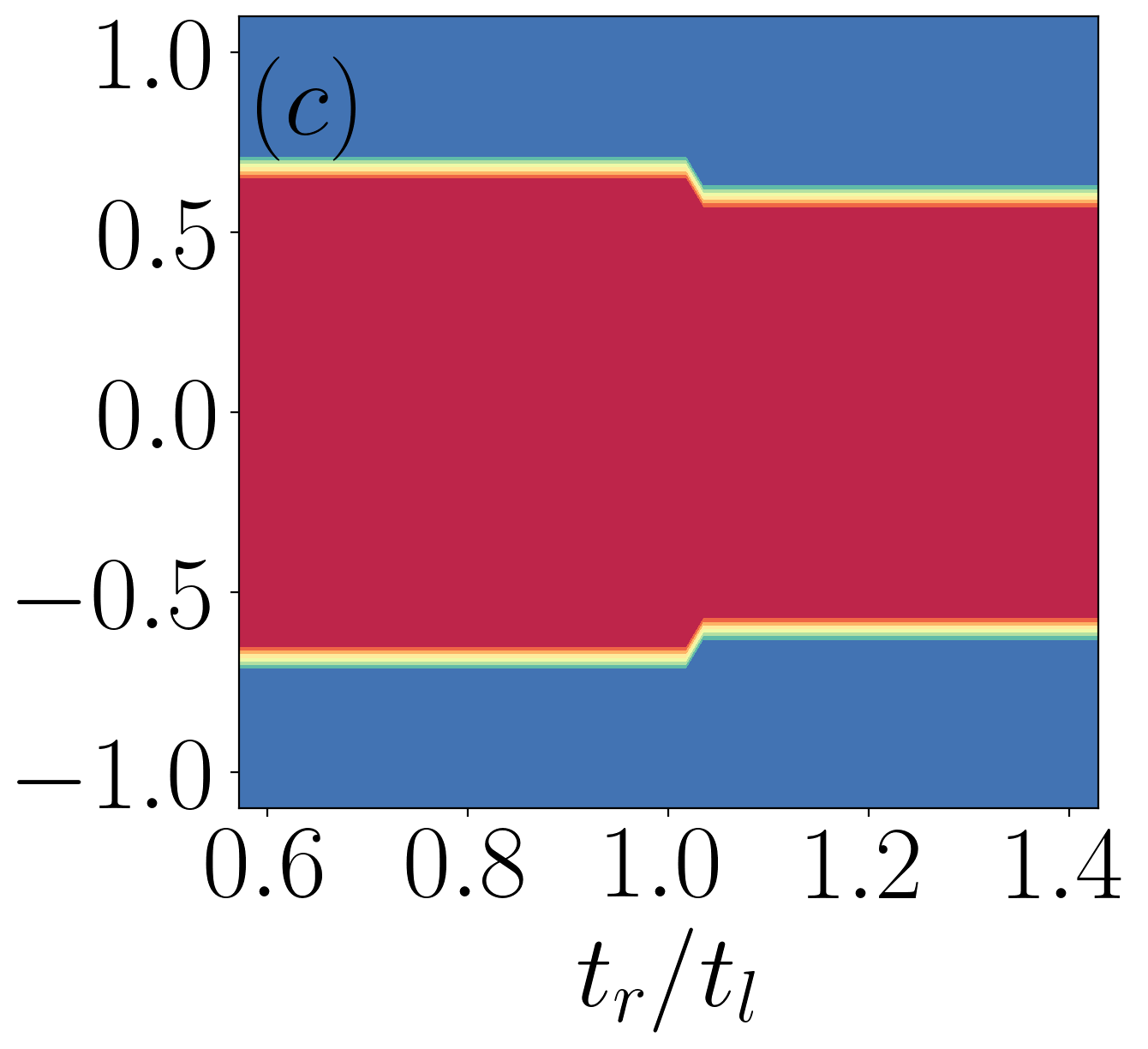}
    \end{subfigure}
        \hfill
    \begin{subfigure}{0.291\columnwidth}
        \centering
        \includegraphics[width=\linewidth]{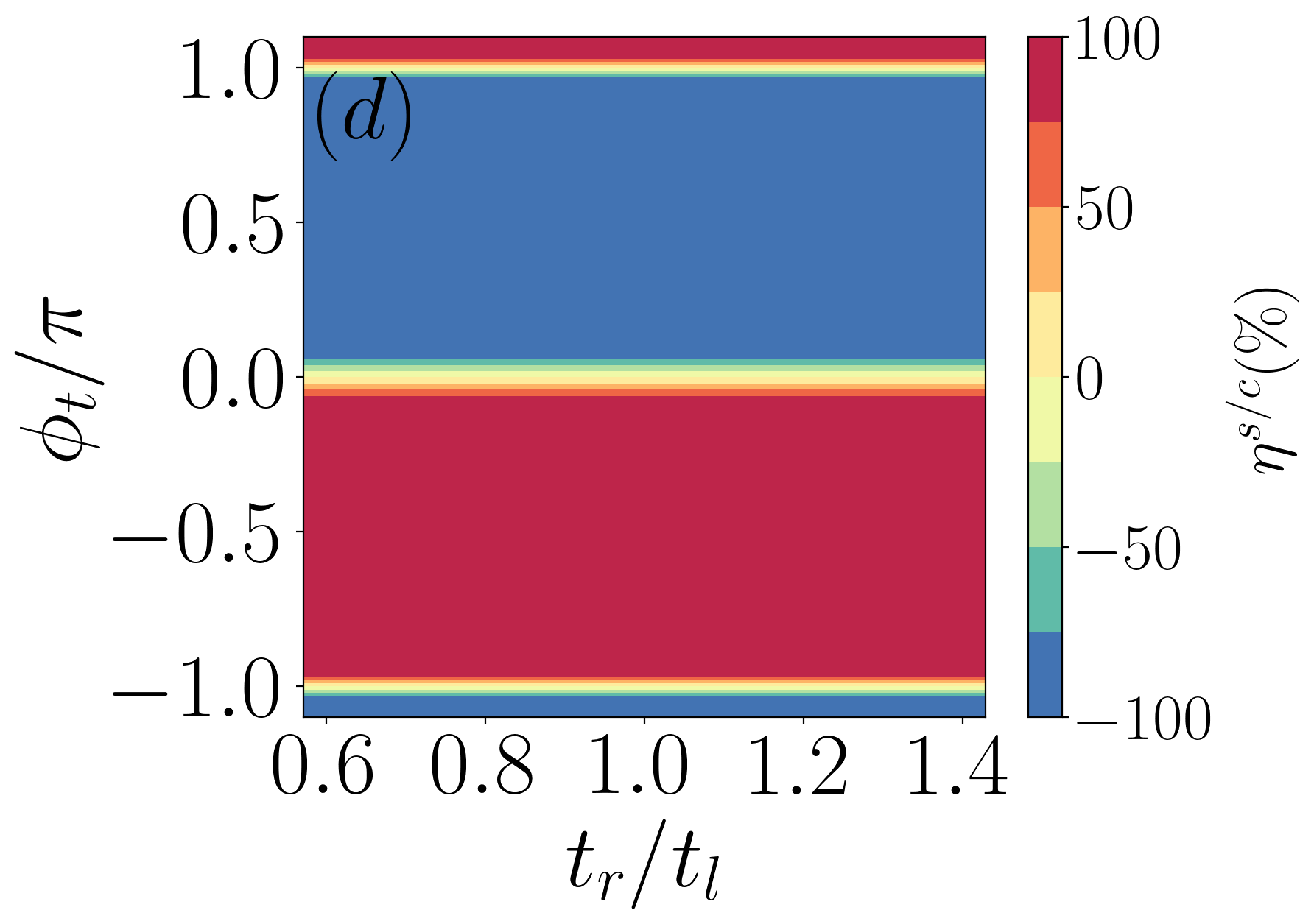}
    \end{subfigure}
    \caption{ Contour plots of the transverse (a,c) SDE and (b,d) CDE, with the color scale representing the diode efficiencies. The horizontal axis corresponds to the interface-coupling ratio $t_{t}/t_b$ in panels (a) and (b) with $t_{r}/t_{l}=1$ and $t_{r}/t_{l}$ in panels (c) and (d) with $t_{t}/t_b=1$, while the vertical axis denotes the transverse phase difference $\phi_{t}$ in all panels. All other system parameters are the same as those used in Fig.~\ref{fig1}.}
    \label{fig5}
\end{figure}

The results presented in the preceding sections, were obtained under idealized conditions, including perfectly transparent interfaces, a fixed junction size, a fixed ratio of pairing amplitudes, zero temperature, and a disorder-free barrier. We now examine the robustness of the $100\%$ diode efficiency by systematically relaxing these assumptions, starting with asymmetric interface couplings. For this, we introduce the coupling amplitude ratios $t_r/t_l$ and $t_t/t_b$ for the left-right and top-bottom interfaces, respectively. Figs.~\ref{fig5}(a) and \ref{fig5}(b) [Figs.~\ref{fig5}(c) and \ref{fig5}(d)] show the contour plots of the transverse charge and spin diode efficiencies, $\eta^{c}$ and $\eta^{s}$, in the $t_t/t_b$ ($t_r/t_l$) versus $\phi_t$ plane. As evident in Figs.~\ref{fig5}(a)--\ref{fig5}(d), the perfect diode effect remains remarkably robust over a broad range of interface asymmetries. We note that the asymmetry breaks both $M_{yz}$ and $M_{xz}$ mirror symmetries, and consequently their combined $\mathcal{W}$ symmetry is also broken. As a result, the current relations in Eqs.~(\ref{eq19}) and (\ref{eq20}) are no longer applicable. Therefore, we adopt the convention defined in Eq.~(\ref{eq21}) to characterize the transverse currents in the presence of asymmetric interface coupling.

\onecolumngrid

\twocolumngrid

\onecolumngrid

\begin{figure}[h]
    \centering

    \begin{subfigure}{0.226\columnwidth}
        \centering
        \includegraphics[width=\linewidth]{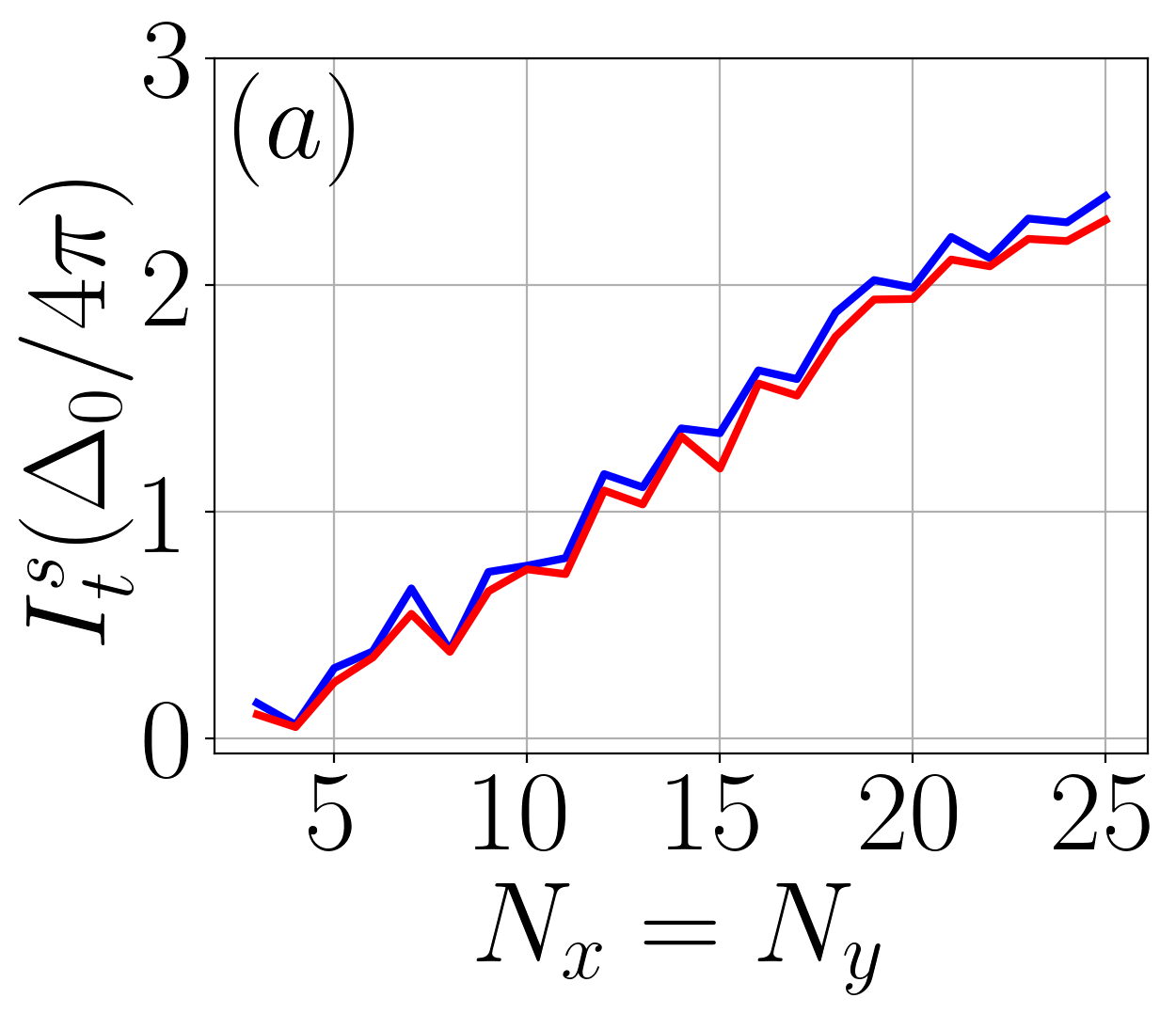}
    \end{subfigure}
    \hfill
    \begin{subfigure}{0.24\columnwidth}
        \centering
        \includegraphics[width=\linewidth]{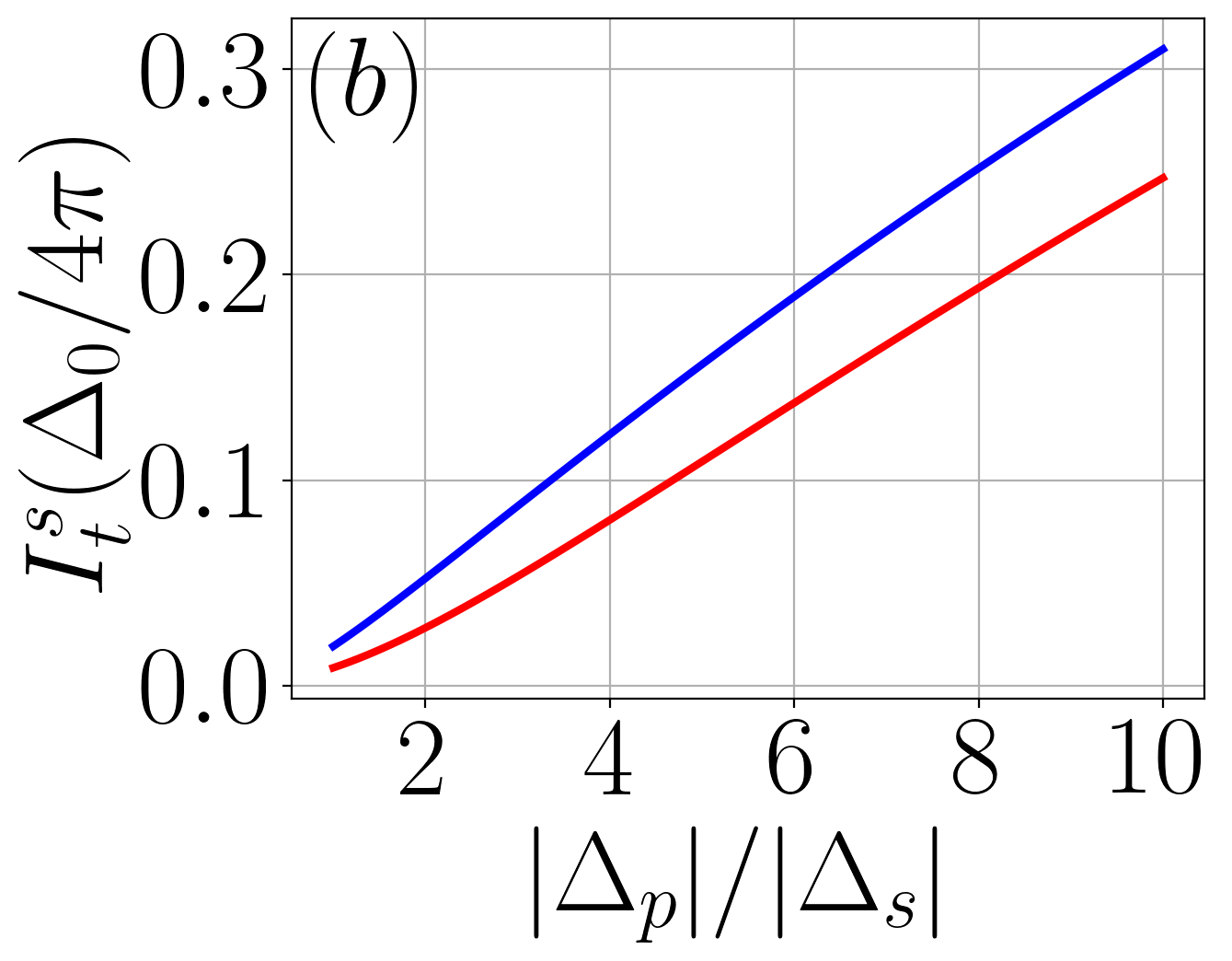}
    \end{subfigure}
    \hfill
    \begin{subfigure}{0.244\columnwidth}
        \centering
        \includegraphics[width=\linewidth]{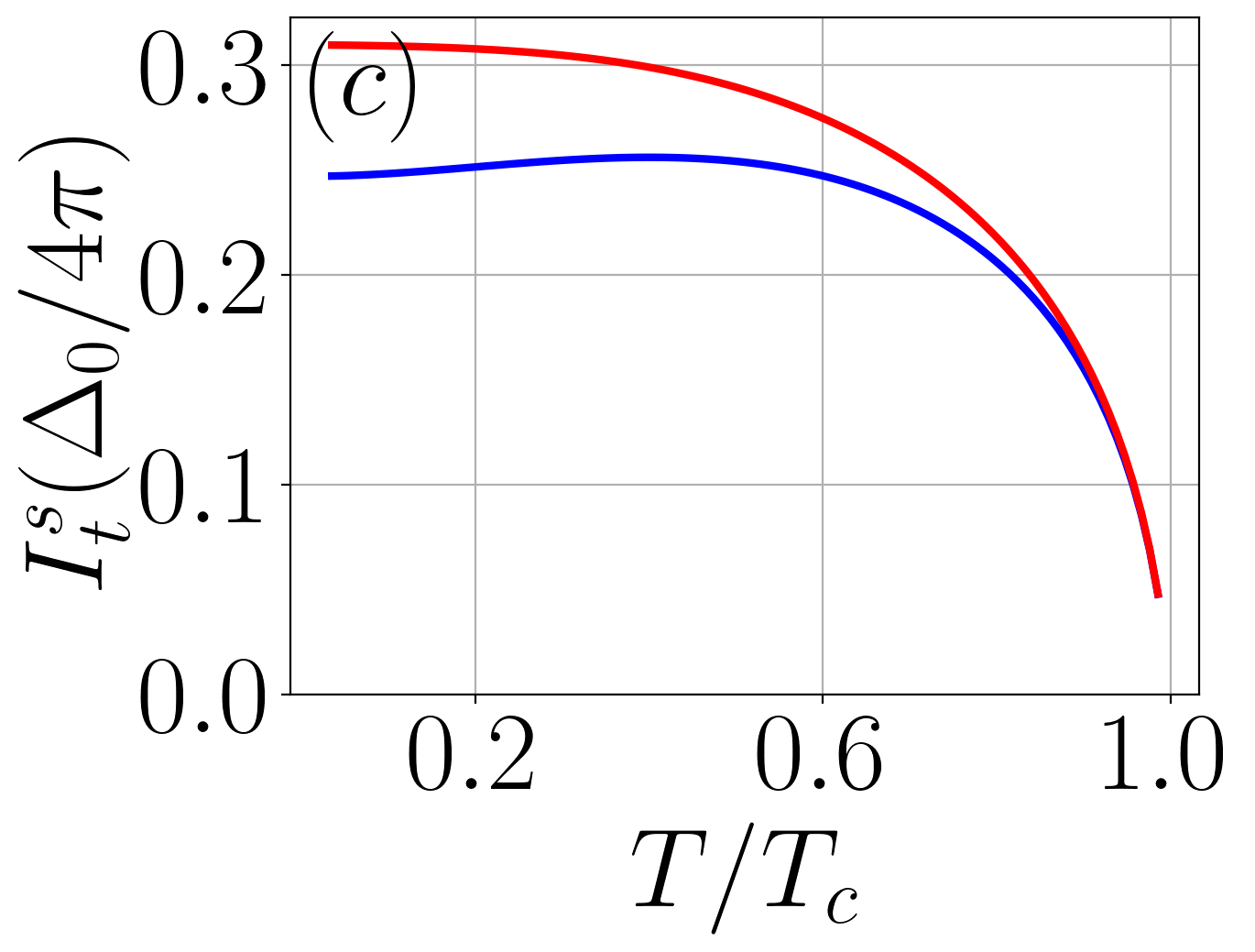}
    \end{subfigure}
        \hfill
    \begin{subfigure}{0.246
    \columnwidth}
        \centering
        \includegraphics[width=\linewidth]{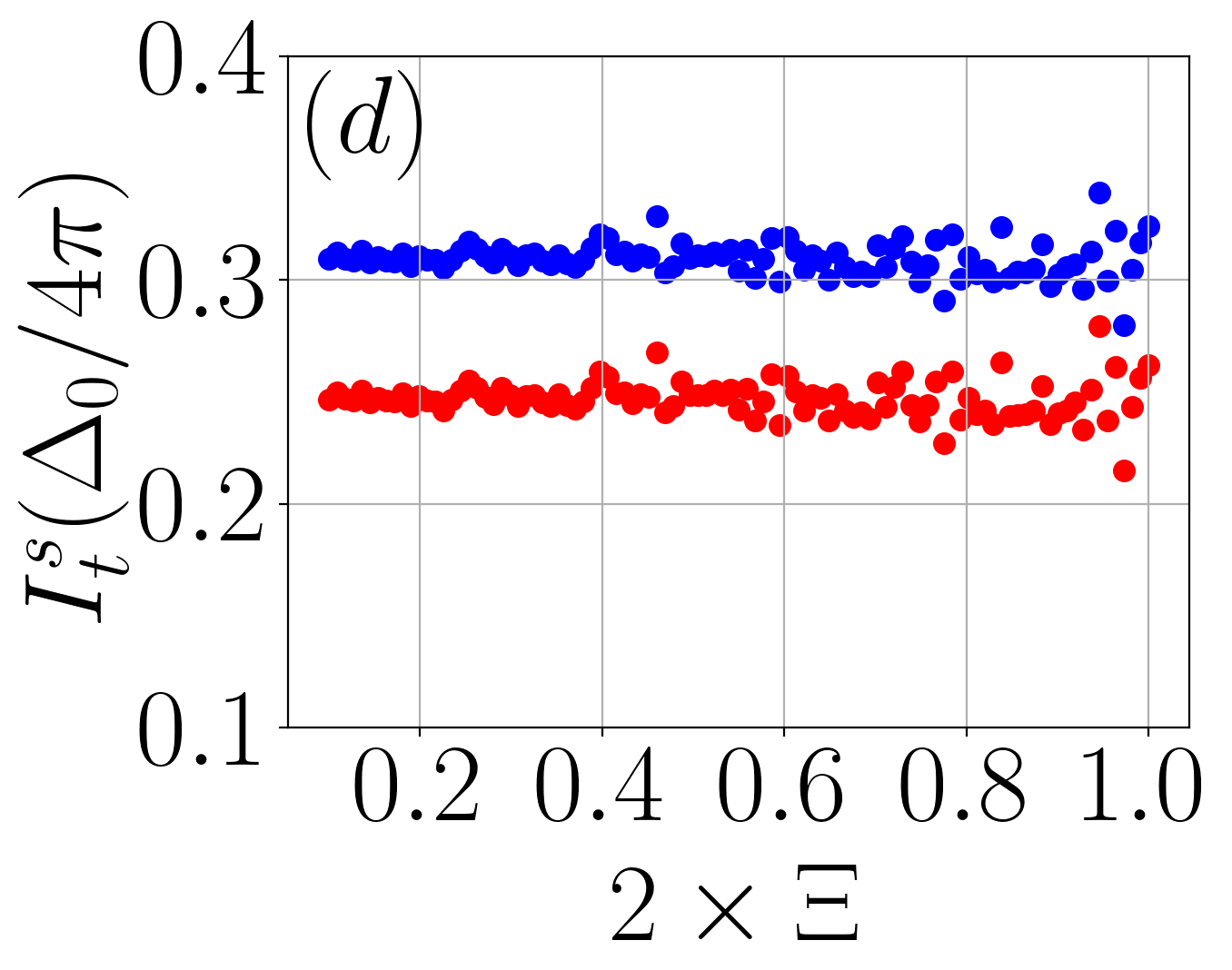}
    \end{subfigure}
    \caption{Maximum (blue) and minimum (red) transverse spin currents over the full range of the longitudinal phase difference, $\phi_{l}$, as functions of (a) the barrier dimensions, $N_x=N_y$, (b) the ratio of the spin-triplet to spin-singlet pairing amplitudes, $|\Delta_p|/|\Delta_s|$, (c) the reduced temperature, $T/T_c$, where $T_c$ is the superconducting critical temperature, and (d) the disorder strength $\Xi$, corresponding to the amplitude of the random on-site potential about $\mu$. Unless otherwise varied along the horizontal axis, all parameters are the same as those used in Fig.~\ref{fig1}, with $\phi_t=0$ and $\alpha=0.2\pi$.}
    \label{fig6}
\end{figure}
\twocolumngrid

In addition, we examine the robustness of the SDE against variations in the system parameters by considering four independent variations, summarized in Fig.~\ref{fig6}. In each panel, we plot the maximum and minimum value of transverse spin currents obtained by varying the longitudinal phase difference $\phi_l$ from $-\pi$ to $\pi$. Although the magnitude of the current changes in response to the different parameters, the diode efficiency remains pinned at its perfect value throughout the parameter ranges considered. First, we vary the junction dimensions by setting $N_x=N_y$, as shown in Fig.~\ref{fig6}(a). The magnitude of the transverse spin current increases with increasing $N_{x/y}$ due to the larger number of transport channels available associated with the increasing junction cross-section. Importantly, both the maximum and the minimum currents retain the same negative sign throughout the entire range of $\phi_l$, thus preserving the perfect diode effect.
However, at larger junction sizes, the increasing separation between the superconducting leads weakens their effective coupling, leading to a competing suppression of the current. 
Moreover, we observe pronounced oscillations with $N_{x/y}$, arising from the discrete quantization of Andreev bound-state in the finite-size mesoscopic junction. Second, we relax the assumption of an order difference in singlet and triplet pairing
amplitudes. Keeping $\Delta_s=0.005 t_0$ fixed, we vary the ratio $|\Delta_p|/|\Delta_s|$ as shown in Fig.~\ref{fig6}(b).
The transverse spin current increases monotonically as the triplet pairing strength is enhanced relative to the singlet component. Nevertheless, both its polarity and the $100\%$ diode efficiency remain unaffected throughout the parameter range, including the case of equal singlet and triplet pairing amplitudes. 
Third, we investigate the effect of temperature, which enters our formalism through two mechanisms: the thermal broadening of the Fermi distribution in Eq.~(\ref{eq7}) and the temperature dependence of the superconducting gap, described by the BCS interpolation $\Delta(T)=\Delta_0\tanh(1.74\sqrt{T_c/T-1})$.  As shown in Fig.~\ref{fig6}(c), the transverse spin current decreases as the temperature increases. This suppression results from the combined effects of thermal occupation of finite-energy states that carry current opposite to those below the Fermi level and the reduction of the superconducting gap. Despite this suppression, the maximum and minimum transverse spin currents remain negative and well separated up to $T/T_c=1$, demonstrating that the perfect diode effect persists throughout the superconducting regime.

Finally, we examine the effect of nonmagnetic disorder by introducing a random on-site potential uniformly distributed within $[-\Xi,+\Xi]$. The transverse spin current is averaged over ten independent disorder realizations for each value of $\Xi$, as shown in Fig.~\ref{fig6}(d). Increasing disorder leads to enhanced fluctuations in the current magnitude; however, its polarity remains unchanged. Remarkably, across all four parameter variations considered in Fig.~\ref{fig6}, the diode efficiency remains essentially fixed at $-100\%$. These results demonstrate that the perfect SDE is not a consequence of fine-tuned parameters but instead reflects a robust property of the junction. Together with its tunability through the gate voltage $V_G$, transverse phase difference $\phi_t$, and crystallographic orientation $\alpha$, this robustness highlights the potential of the proposed junction for dissipationless spin-transport applications.
\vspace{0.5cm}
\section{Conclusion}
In this work, we have studied the transverse spin and charge transport in a four-terminal JJ with NM barrier. The left and right leads are PMs with proximity-induced conventional $s$-wave superconductivity, while the top and bottom leads are spin-triplet $p$-wave SCs. When the transverse phase difference is zero, the junction supports a pure transverse spin current whose dependence on the longitudinal phase exhibits a perfect $100\%$ SDE. Remarkably, this perfect spin nonreciprocity persists for finite transverse phase differences, where it is accompanied by a perfect $100\%$ CDE. Both the transverse SDE and CDE exhibit step-like variations with the transverse phase. We further demonstrate that the diode polarity can be electrically controlled through a gate potential applied to the normal region. In particular, the transverse SDE undergoes a sharp transition between $+100\%$ and $-100\%$, enabling complete reversal of the diode polarity. The SDE also exhibits an exact step-like dependence on the crystallographic orientation of the PM, reflecting the reversal of the spin splitting as the magnetic lobe orientation crosses the $x$ axis. In contrast, the CDE remains largely insensitive to the crystallographic orientation, providing independent control of spin and charge rectification. Furthermore, the perfect diode effect persists against asymmetric interface couplings, variations in the pairing amplitudes, temperature, nonmagnetic disorder, and junction dimensions, demonstrating that it does not rely on fine-tuned parameters.

Our results establish this four-terminal JJ as a robust and highly tunable platform for phase-controlled nonreciprocal spin and charge transport. The combination of gate voltage, transverse superconducting phase, and crystallographic orientation provides complementary electrical and structural knobs for controlling the magnitude and polarity of the diode response, including full-efficiency rectification and complete reversal of the diode polarity. These features highlight the potential of this architecture for dissipationless and reconfigurable superconducting spintronics applications.
\label{sec6}

\section{ACKNOWLEDGMENTS}
For financial support, L.S. thanks UGC, India, B. G. thanks DST, India, and M.T. gratefully acknowledges  Indian Institute of Technology Hyderabad, India. The authors thank A. Soori, Y. Fukaya and Y. Tanaka, for stimulating discussions on related topics.

%\section{DATA AVAILABILITY}
%The data supporting the findings of this article are not publicly available. The data are available from the authors upon reasonable request.

\hspace{1cm}

\onecolumngrid
\appendix

\section{Matrix Representation of Hamiltonian}

In this section, we present the matrix representation of the tight-binding 
Hamiltonian given in Eq.~(\ref{eq2}--\ref{eq5}) in the Nambu basis. In the basis $\psi=(c_{i,j,\uparrow}, c_{i,j,\downarrow}, c^{\dagger}_{i,j,\uparrow}, c^{\dagger}_{i,j,\downarrow})$, the matrices are, 
\begin{align}
    H_{0}^{\nu,\kappa} &= (4t_{0}-\mu)\tau_{z}\otimes\sigma_{0} - \Delta_{s}\big(\tau_{y}\cos{(\phi_\nu/2)} - \tau_{x}\sin{(\phi_\nu/2)}\big)\otimes\sigma_{y} + \kappa J_{sd}\tau_{z}\otimes\sigma_{x},\\
    H_{x}^{\nu}&=-t_{0}\tau_{z}\otimes\sigma_{0} + \frac{t_{j}}{2i}\cos{\alpha}\tau_{0}\otimes\sigma_{z},\\
     H_{y}^{\nu}&=-t_{0}\tau_{z}\otimes\sigma_{0} + \frac{t_{j}}{2i}\sin{\alpha}\tau_{0}\otimes\sigma_{z},\\
    H_{0}^{N}&=(4t_{0}-\mu)\tau_{z}\otimes\sigma_{0},\\
    H_{x/y}^{N}&=-t_{0}\tau_{z}\otimes\sigma_{0},\\
    H_{0}^{\mu}&=(4t_{0}-\mu)\tau_{z}\otimes\sigma_{0},\\
    H_{x}^{\mu}&=-t_{0}\tau_{z}\otimes\sigma_{0},\\
    H_{y}^{\mu}&=-t_{0}\tau_{z}\otimes\sigma_{0} - \frac{\Delta_{p}}{2i}\big(  \cos{(\phi_\mu/2)}\tau_{0}  + \sin{(\phi_\mu/2)} \tau_{y}\big)\otimes\sigma_{z}.
\end{align}

Here, $H_{0}$ represents the onsite matrix, and $H_{x}$ ($H_{y}$) represents the hopping matrix along the $x$-direction ($y$-direction). For the calculation of the self-energy and surface Green's function of 
each lead, we require the matrix representation of the Hamiltonian of 
vertical (horizontal) strips of the lattice in the left and right 
(top and bottom) leads. For that case we introduce,
\begin{align}
    H_{11}^{\nu,\kappa(\mu)} &= \begin{bmatrix}
        H_{0}^{\nu,\kappa(\mu)}  & {H_{y(x)}^{\nu(\mu)}}^{\dagger} &  & & \\
        {H_{y(x)}^{\nu(\mu)}} & H_{0}^{\nu,\kappa(\mu)} &{H_{y(x)}^{\nu(\mu)}}^{\dagger} & &\\
        &  {H_{y(x)}^{\nu(\mu)}}&H_{0}^{\nu,\kappa(\mu)} & &\\
         & & ... & ... & {H_{y(x)}^{\nu(\mu)}}^{\dagger}\\
         &&& {H_{y(x)}^{\nu(\mu)}}&H_{0}^{\nu(\mu)}
    \end{bmatrix},\\
    H_{12}^{\nu (\mu)} &= \begin{bmatrix}
        H_{x(y)}^{\nu(\mu)} & & & & \\
        & H_{x(y)}^{\nu(\mu)} & & &\\
        & & H_{x(y)}^{\nu(\mu)} & &\\
        &&&...&\\
        &&&& H_{x(y)}^{\nu(\mu)}
    \end{bmatrix},
\end{align}

where $H_{11}^{\nu(\mu)}$ is the on-strip block Hamiltonian of the 
vertical (horizontal) strips 
and $H_{12}^{\nu(\mu)}$ is the matrix representing the hopping along 
the $+x$ ($+y$) direction in the in the $\nu$=left/right ($\mu=$ top/bottom) leads, respectively.

\label{AppA}

\section{Surface and non-local Green's function}

\label{AppB}

As shown in Eq.~(\ref{eq8}), to calculate the current we require the 
retarded and advanced surface Green's functions of each isolated 
semi-infinite lead, which are obtained using the M\"{o}bius 
transformation [\onlinecite{PhysRevB.55.5266}]. To this end, we first define M\"{o}bius transformation matrix for each lead,

\begin{align}
    X_{L(B)} &= \begin{bmatrix}
        0 &{H_{12}^{L(B)}}^{-1}\\
        -{H_{12}^{L(B)}}^{\dagger} & \big((E+i\zeta) - H_{11}^{L(B)}\big){H_{12}^{L(B)}}^{-1}
    \end{bmatrix},\\
    X_{R(T)} &= \begin{bmatrix}
        0 &{\big( {H_{12}^{R(T)} }^{\dagger}\big)}^{-1}\\
        -{H_{12}^{R(T)}}& \big((E+i\zeta) - H_{11}^{R(T)}\big){\big( {H_{12}^{R(T)} }^{\dagger}\big)}^{-1}
    \end{bmatrix}.
\end{align}

Here, we have omitted the $\kappa$ index, since the formalism is the same and separable. We then diagonalize each of these transformation matrices using $U_{\beta}$, such that $U_\beta^{-1} X_{\beta}U_{\beta}=\Lambda_{\beta}$. Here $\Lambda_{\beta}$ is a diagonal matrix consisting of eigenvalues in ascending order. Then the surface Green's function for each lead is given by $g_{\beta}^{r}= U_{\beta}^{12} {U_{\beta}^{22}}^{-1}$, where
\begin{align}
    U_{\beta}= \begin{bmatrix}
        U^{11}_{\beta} & U^{12}_{\beta}\\
        U^{21}_{\beta} & U^{22}_{\beta}
    \end{bmatrix}.
\end{align}

The retarded Green's function of the normal region is given by,
\begin{align}
    G_{N}^{r} = \left[E + i\zeta - H_{N} - \Sigma_{L} - \Sigma_{R} 
    - \Sigma_{T} - \Sigma_{B}\right]^{-1},
\end{align}
where $H_{N}$ is the BdG Hamiltonian of the normal barrier region, 
$\zeta \to 0^{+}$ is a positive infinitesimal broadening factor, and 
$\Sigma_{\beta}={\hat{T}_{\beta}}^{\dagger}g^{r}_{\beta}\hat{T}_{\beta}$ 
is the retarded self-energy due to lead $\beta$. The surface Green's function $G_{NN}^{r}$ for each lead $\beta$ is 
extracted as the submatrix of $G_{N}^{r}$ corresponding to the 
edge lattice sites adjacent to the interface with lead $\beta$.

\section{Matrix Representation of Transformation Operator}
\label{AppC}
In this section we present the matrix representation of Transformation operator defined in Eq.(\ref{eq10}). These include the time reversal $\mathcal{T}$, mirror in $xz$ and $yz$-plane $M_{xz}$ and $M_{yz}$, and spin rotation of $180^{\circ}$ about $x$, $y$ and $z$-axis, $R_{x}$, $R_{y}$ and $R_{z}$. The matrix representations are as follows,

\begin{align}
    \mathcal{T}&=\begin{bmatrix}
        -i \sigma_y &0\\
        0 & -i\sigma_y
    \end{bmatrix}\mathcal{K},\\
    M_{xz}&=\begin{bmatrix}
        i\sigma_{y} & 0 \\
        0  &  i \sigma_y
    \end{bmatrix}\mathcal{P}_y,\\
    M_{yz}&=\begin{bmatrix}
        i\sigma_x & 0\\
        0 & -i\sigma_x
    \end{bmatrix}\mathcal{P}_{x},\\
    R_{y}&=\begin{bmatrix}
        -i\sigma_y & 0\\
        0 & -i\sigma_y
    \end{bmatrix},\\
    R_{z}&=\begin{bmatrix}
        -i\sigma_z & 0\\
        0 & i\sigma_z
    \end{bmatrix},\\
     R_{x}&=\begin{bmatrix}
        -i\sigma_x & 0\\
        0 & i\sigma_x
    \end{bmatrix}.
\end{align}
In each case, the upper (lower) block acts on the particle (hole) sector, and the
hole block is the complex conjugate of the particle block. Here $\mathcal{K}$
denotes complex conjugation and $\mathcal{P}_{x(y)}$ the spatial inversion
$x\to-x$ ($y\to-y$). None of these operators acts on the sectoral index, the
mapping $H^{\kappa}\to H^{\bar{\kappa}}$ discussed in Sec.~\ref{sec3} arises
instead from the reversal of $\sigma_x$ in the $sd$ exchange term. The transformation of some important terms of the Hamiltonian in Eq.(\ref{eq2}-\ref{eq5}) is given in Table.[\ref{tab}]

\begin{table}[H]
    \centering
    \begin{tabular}{|c|c|c|c|c|c|c|}
    \hline
        \diagbox{Terms}{Operator}   & $\mathcal{T}$ & $M_{xz}$ &  $M_{yz}$ & $R_x$ & $R_z$ & $R_{y}$\\ \hline
        $PM(\alpha=0)$ & $+$ & $-$ & $+$& $-$&$+$ &$-$\\ \hline
        $PM(\alpha=\pi/2)$ & $+$ & $+$&$-$ & $-$&$+$& $-$\\ \hline
        $\mathbf{\Delta_{s}}(\phi)$ & $\phi\rightarrow-\phi$ & $+$& $+$& $+$& $+$ &$+$\\ \hline
        $\mathbf{\Delta_{p}}(\phi)$ & $\phi\rightarrow-\phi$ & $+$& $+$& $+$&$\phi\rightarrow\phi\pm\pi$ &$\phi \rightarrow \phi\pm\pi$\\ \hline
        $J_{sd}$ &$-$ &$-$ & $+$ & $-$ & $-$& $+$\\ \hline
    \end{tabular}
    \caption{Transformation of the terms given in Eq.~(\ref{eq2}-\ref{eq5}), where $\mathbf{\Delta_{s}}$ is pairing potential of spin singlet $s-$wave superconductivity and $\mathbf{\Delta_{p}}$ is pairing potential of equal-spin spin triplet $p_y-$wave superconductivity.}
    \label{tab}
\end{table}

\twocolumngrid
\bibliography{biblo}

@ARTICLE{6773080,
  author={Shockley, W.},
  journal={The Bell System Technical Journal}, 
  title={The theory of p-n junctions in semiconductors and p-n junction transistors}, 
  year={1949},
  volume={28},
  number={3},
  pages={435-489},
  doi={10.1002/j.1538-7305.1949.tb03645.x}}

@ARTICLE{6372252,
  author={Losco, E. F.},
  journal={Transactions of the American Institute of Electrical Engineers, Part I: Communication and Electronics}, 
  title={Properties of silicon power rectifiers}, 
  year={1955},
  volume={74},
  number={1},
  pages={106-111},
  doi={10.1109/TCE.1955.6372252}}

@Article{Ando2020,
author={Ando, F.
and Miyasaka, Y.
and Li, T.
and Ishizuka, J.
and Arakawa, T.
and Shiota, Y.
and Moriyama, T.
and Yanase, Y.
and Ono, T.},
title={Observation of superconducting diode effect},
journal={Nature},
year={2020},
month={Aug},
day={01},
volume={584},
number={7821},
pages={373-376},
issn={1476-4687},
doi={10.1038/s41586-020-2590-4},
url={https://doi.org/10.1038/s41586-020-2590-4}
}

@article{PhysRevB.49.9244,
  title = {Asymmetric current-voltage characteristics in type-II superconductors},
  author = {Jiang, Xiuguang and Connolly, P. J. and Hagen, S. J. and Lobb, C. J.},
  journal = {Phys. Rev. B},
  volume = {49},
  issue = {13},
  pages = {9244--9247},
  numpages = {0},
  year = {1994},
  month = {Apr},
  publisher = {American Physical Society},
  doi = {10.1103/PhysRevB.49.9244},
  url = {https://link.aps.org/doi/10.1103/PhysRevB.49.9244}
}

@article{Miyasaka_2021,
doi = {10.35848/1882-0786/ac03c0},
url = {https://dx.doi.org/10.35848/1882-0786/ac03c0},
year = {2021},
month = {jun},
publisher = {IOP Publishing},
volume = {14},
number = {7},
pages = {073003},
author = {Miyasaka, Yuta and Kawarazaki, Ryo and Narita, Hideki and Ando, Fuyuki and Ikeda, Yuhei and Hisatomi, Ryusuke and Daido, Akito and Shiota, Yoichi and Moriyama, Takahiro and Yanase, Youichi and Ono, Teruo},
title = {Observation of nonreciprocal superconducting critical field},
journal = {Applied Physics Express}
}

@article{PhysRevB.107.224518,
  title = {Superconducting diode effect in quasi-one-dimensional systems},
  author = {de Picoli, Tatiana and Blood, Zane and Lyanda-Geller, Yuli and V\"ayrynen, Jukka I.},
  journal = {Phys. Rev. B},
  volume = {107},
  issue = {22},
  pages = {224518},
  numpages = {6},
  year = {2023},
  month = {Jun},
  publisher = {American Physical Society},
  doi = {10.1103/PhysRevB.107.224518},
  url = {https://link.aps.org/doi/10.1103/PhysRevB.107.224518}
}

@Article{Coraiola2024,
author={Coraiola, Marco
and Svetogorov, Aleksandr E.
and Haxell, Daniel Z.
and Sabonis, Deividas
and Hinderling, Manuel
and ten Kate, Sofieke C.
and Cheah, Erik
and Krizek, Filip
and Schott, R{\"u}diger
and Wegscheider, Werner
and Cuevas, Juan Carlos
and Belzig, Wolfgang
and Nichele, Fabrizio},
title={Flux-Tunable Josephson Diode Effect in a Hybrid Four-Terminal Josephson Junction},
journal={ACS Nano},
year={2024},
month={Mar},
day={26},
publisher={American Chemical Society},
volume={18},
number={12},
pages={9221-9231},
issn={1936-0851},
doi={10.1021/acsnano.4c01642},
url={https://doi.org/10.1021/acsnano.4c01642}
}

@article{Gupta2023,
  author  = {Gupta, Mohit and Graziano, Gian V. and Pendharkar, Mihir and others},
  title   = {Gate-tunable superconducting diode effect in a three-terminal Josephson device},
  journal = {\href{https://doi.org/10.1038/s41467-023-38856-0}{Nat. Commun. \textbf{14}, 3078 (2023)}},
}

@Article{10.21468/SciPostPhys.17.2.037,
	title={{Theory of universal diode effect in three-terminal Josephson junctions}},
	author={Jorge Huamani Correa and Michał P. Nowak},
	journal={SciPost Phys.},
	volume={17},
	pages={037},
	year={2024},
	publisher={SciPost},
	doi={10.21468/SciPostPhys.17.2.037},
	url={https://scipost.org/10.21468/SciPostPhys.17.2.037},
}

@article{Sahoo_2025,
doi = {10.1088/1361-648X/adf1d0},
url = {https://doi.org/10.1088/1361-648X/adf1d0},
year = {2025},
month = {jul},
publisher = {IOP Publishing},
volume = {37},
number = {30},
pages = {305302},
author = {Sahoo, Bijay Kumar and Soori, Abhiram},
title = {Four-terminal Josephson junctions: diode effects, anomalous currents and transverse currents},
journal = {Journal of Physics: Condensed Matter}
}

@misc{sahoo2026giantfieldfreetransversejosephson,
      title={Giant field-free transverse Josephson diode effect in altermagnets}, 
      author={Bijay Kumar Sahoo and Abhiram Soori},
      year={2025},
      eprint={2509.14109},
      archivePrefix={arXiv},
      primaryClass={cond-mat.mes-hall}, 
}

@misc{Sibgat,
      title={Pure Spin Bulk Photovoltaic Effect in an Altermagnetic Higher-Order Topological Insulator}, 
      author={Sibgat Ulah and Ankan Bhattacharyya and Manisha Thakurathi},
      year={2026},
      eprint={2607.19018},
      archivePrefix={arXiv},
      primaryClass={cond-mat.mes-hall}, 
}

@article{Bao2013,
  author  = {Bao, Zhi-Qiang and Xie, X. C. and Sun, Qing-Feng},
  title   = {Ginzburg--Landau-type theory of spin superconductivity},
  journal = {\href{https://doi.org/10.1038/ncomms3951}{Nat. Commun. \textbf{4}, 2951 (2013)}},
}

@article{PhysRevB.105.184511,
  title = {Spin phase regulated spin Josephson supercurrent in topological superconductor},
  author = {Mao, Yue and Sun, Qing-Feng},
  journal = {Phys. Rev. B},
  volume = {105},
  issue = {18},
  pages = {184511},
  numpages = {8},
  year = {2022},
  month = {May},
  publisher = {American Physical Society},
  doi = {10.1103/PhysRevB.105.184511},
  url = {https://link.aps.org/doi/10.1103/PhysRevB.105.184511}
}

@article{PhysRevB.90.155450,
  title = {Single fermion manipulation via superconducting phase differences in multiterminal Josephson junctions},
  author = {van Heck, B. and Mi, S. and Akhmerov, A. R.},
  journal = {Phys. Rev. B},
  volume = {90},
  issue = {15},
  pages = {155450},
  numpages = {9},
  year = {2014},
  month = {Oct},
  publisher = {American Physical Society},
  doi = {10.1103/PhysRevB.90.155450},
  url = {https://link.aps.org/doi/10.1103/PhysRevB.90.155450}
}

@article{PhysRevResearch.2.023197,
  title = {Transport signatures of bulk topological phases in double Rashba nanowires probed by spin-polarized STM},
  author = {Thakurathi, Manisha and Chevallier, Denis and Loss, Daniel and Klinovaja, Jelena},
  journal = {Phys. Rev. Res.},
  volume = {2},
  issue = {2},
  pages = {023197},
  numpages = {12},
  year = {2020},
  month = {May},
  publisher = {American Physical Society},
  doi = {10.1103/PhysRevResearch.2.023197},
  url = {https://link.aps.org/doi/10.1103/PhysRevResearch.2.023197}
}

@article{Riwar2016,
  author  = {Riwar, Roman-Pascal and Houzet, Manuel and Meyer, Julia S. and Nazarov, Yuli V.},
  title   = {Multi-terminal josephson junctions as topological matter},
  journal = {\href{https://doi.org/10.1038/ncomms11167}{Nat.  Commun. \textbf{7}, 11167 (2016)}},
}

@article{h2qg-qhf7,
  title = {Voltage-tunable spin supercurrent nonreciprocity reaching 100\% efficiency},
  author = {Sun, Chi and Tjernshaugen, Johanne Bratland and Linder, Jacob},
  journal = {Phys. Rev. B},
  volume = {112},
  issue = {6},
  pages = {064504},
  numpages = {10},
  year = {2025},
  month = {Aug},
  publisher = {American Physical Society},
  doi = {10.1103/h2qg-qhf7},
  url = {https://link.aps.org/doi/10.1103/h2qg-qhf7}
}

@misc{fu2026perfectspinnonreciprocitygated,
      title={Perfect spin nonreciprocity in gated superconducting altermagnetic heterostructures}, 
      author={Pei-Hao Fu and Jun-Feng Liu and Luca Chirolli and Jorge Cayao},
      year={2026},
      eprint={2604.20312},
      archivePrefix={arXiv},
      primaryClass={cond-mat.supr-con}, 
}

@article{4t18-yyx4,
  title = {Theory of quantum-geometric charge and spin Josephson diode effects in strongly spin-polarized hybrid structures with noncoplanar spin textures},
  author = {Schulz, Niklas L. and Nikoli\ifmmode \acute{c}\else \'{c}\fi{}, Danilo and Eschrig, Matthias},
  journal = {Phys. Rev. B},
  volume = {112},
  issue = {10},
  pages = {104515},
  numpages = {24},
  year = {2025},
  month = {Sep},
  publisher = {American Physical Society},
  doi = {10.1103/4t18-yyx4},
  url = {https://link.aps.org/doi/10.1103/4t18-yyx4}
}

@article{nb38-v1jq,
  title = {Quantum-geometric spin and charge Josephson diode effects},
  author = {Schulz, Niklas L. and Nikoli\ifmmode \acute{c}\else \'{c}\fi{}, Danilo and Eschrig, Matthias},
  journal = {Phys. Rev. B},
  volume = {112},
  issue = {10},
  pages = {104514},
  numpages = {6},
  year = {2025},
  month = {Sep},
  publisher = {American Physical Society},
  doi = {10.1103/nb38-v1jq},
  url = {https://link.aps.org/doi/10.1103/nb38-v1jq}
}

@misc{nikolic2025necessaryconditionsspinresolvedjosephson,
      title={Necessary conditions for spin-resolved Josephson diode effect across strongly spin-polarized magnetic materials}, 
      author={Danilo Nikolić and Niklas L. Schulz and Matthias Eschrig},
      year={2025},
      eprint={2512.22017},
      archivePrefix={arXiv},
      primaryClass={cond-mat.supr-con}, 
}

@article{PhysRevLett.132.216001,
  title = {Universal Spin Superconducting Diode Effect from Spin-Orbit Coupling},
  author = {Mao, Yue and Yan, Qing and Zhuang, Yu-Chen and Sun, Qing-Feng},
  journal = {Phys. Rev. Lett.},
  volume = {132},
  issue = {21},
  pages = {216001},
  numpages = {7},
  year = {2024},
  month = {May},
  publisher = {American Physical Society},
  doi = {10.1103/PhysRevLett.132.216001},
  url = {https://link.aps.org/doi/10.1103/PhysRevLett.132.216001}
}

@article{Yang_2026,
doi = {10.1209/0295-5075/ae488f},
url = {https://doi.org/10.1209/0295-5075/ae488f},
year = {2026},
month = {mar},
publisher = {EDP Sciences, IOP Publishing and Società Italiana di Fisica},
volume = {153},
number = {5},
pages = {56001},
author = {Yang, Zhe and Niu, Zhi Ping and Liu, Xin},
title = {Perfect spin Josephson diode effect in unconventional p-wave magnet},
journal = {Europhysics Letters}
}

@article{Eschrig_2015,
doi = {10.1088/0034-4885/78/10/104501},
url = {https://doi.org/10.1088/0034-4885/78/10/104501},
year = {2015},
month = {sep},
publisher = {IOP Publishing},
volume = {78},
number = {10},
pages = {104501},
author = {Eschrig, Matthias},
title = {Spin-polarized supercurrents for spintronics: a review of current progress},
journal = {Reports on Progress in Physics}
}

@article{Linder2015,
author={Linder, Jacob
and Robinson, Jason W. A.},
title={Superconducting spintronics},
journal={Nat. Phys.},
year={2015},
month={Apr},
day={01},
volume={11},
number={4},
pages={307-315},
issn={1745-2481},
doi={10.1038/nphys3242},
url={https://doi.org/10.1038/nphys3242}
}

@article{RevModPhys.76.323,
  title = {Spintronics: Fundamentals and applications},
  author = {\ifmmode \check{Z}\else \v{Z}\fi{}uti\ifmmode \acute{c}\else \'{c}\fi{}, Igor and Fabian, Jaroslav and Das Sarma, S.},
  journal = {Rev. Mod. Phys.},
  volume = {76},
  issue = {2},
  pages = {323--410},
  numpages = {0},
  year = {2004},
  month = {Apr},
  publisher = {American Physical Society},
  doi = {10.1103/RevModPhys.76.323},
  url = {https://link.aps.org/doi/10.1103/RevModPhys.76.323}
}

@article{RevModPhys.80.1517,
  title = {Nobel Lecture: Origin, development, and future of spintronics},
  author = {Fert, Albert},
  journal = {Rev. Mod. Phys.},
  volume = {80},
  issue = {4},
  pages = {1517--1530},
  numpages = {0},
  year = {2008},
  month = {Dec},
  publisher = {American Physical Society},
  doi = {10.1103/RevModPhys.80.1517},
  url = {https://link.aps.org/doi/10.1103/RevModPhys.80.1517}
}

@article{PhysRevB.104.134514,
  title = {Spin-triplet superconductor--quantum anomalous Hall insulator--spin-triplet superconductor Josephson junctions: $0\text{\ensuremath{-}}\ensuremath{\pi}$ transition, ${\ensuremath{\phi}}_{0}$ phase, and switching effects},
  author = {Cheng, Qiang and Yan, Qing and Sun, Qing-Feng},
  journal = {Phys. Rev. B},
  volume = {104},
  issue = {13},
  pages = {134514},
  numpages = {9},
  year = {2021},
  month = {Oct},
  publisher = {American Physical Society},
  doi = {10.1103/PhysRevB.104.134514},
  url = {https://link.aps.org/doi/10.1103/PhysRevB.104.134514}
}

@article{PhysRevB.93.195302,
  title = {Quantum interference in topological insulator Josephson junctions},
  author = {Song, Juntao and Liu, Haiwen and Liu, Jie and Li, Yu-Xian and Joynt, Robert and Sun, Qing-feng and Xie, X. C.},
  journal = {Phys. Rev. B},
  volume = {93},
  issue = {19},
  pages = {195302},
  numpages = {8},
  year = {2016},
  month = {May},
  publisher = {American Physical Society},
  doi = {10.1103/PhysRevB.93.195302},
  url = {https://link.aps.org/doi/10.1103/PhysRevB.93.195302}
}

@article{PhysRevB.55.5266,
  title = {Closed-form solutions to surface Green's functions},
  author = {Umerski, A.},
  journal = {Phys. Rev. B},
  volume = {55},
  issue = {8},
  pages = {5266--5275},
  numpages = {0},
  year = {1997},
  month = {Feb},
  publisher = {American Physical Society},
  doi = {10.1103/PhysRevB.55.5266},
  url = {https://link.aps.org/doi/10.1103/PhysRevB.55.5266}
}

@article{Sun_2009,
doi = {10.1088/0953-8984/21/34/344204},
url = {https://doi.org/10.1088/0953-8984/21/34/344204},
year = {2009},
month = {jul},
publisher = {},
volume = {21},
number = {34},
pages = {344204},
author = {Sun, Qing-feng and Xie, X C},
title = {Quantum transport through a graphene nanoribbon–superconductor junction},
journal = {Journal of Physics: Condensed Matter}
}

@article{yqsg-xdg8,
  title = {Tunable Josephson diode effect in singlet superconductor-altermagnet-triplet superconductor junctions},
  author = {Sharma, Lovy and Thakurathi, Manisha},
  journal = {Phys. Rev. B},
  volume = {112},
  issue = {10},
  pages = {104506},
  numpages = {10},
  year = {2025},
  month = {Sep},
  publisher = {American Physical Society},
  doi = {10.1103/yqsg-xdg8},
  url = {https://link.aps.org/doi/10.1103/yqsg-xdg8}
}

@article{PhysRevB.106.214524,
  title = {Theory of giant diode effect in $d$-wave superconductor junctions on the surface of a topological insulator},
  author = {Tanaka, Yukio and Lu, Bo and Nagaosa, Naoto},
  journal = {Phys. Rev. B},
  volume = {106},
  issue = {21},
  pages = {214524},
  numpages = {13},
  year = {2022},
  month = {Dec},
  publisher = {American Physical Society},
  doi = {10.1103/PhysRevB.106.214524},
  url = {https://link.aps.org/doi/10.1103/PhysRevB.106.214524}
}

@article{PhysRevLett.131.096001,
  title = {Tunable Josephson Diode Effect on the Surface of Topological Insulators},
  author = {Lu, Bo and Ikegaya, Satoshi and Burset, Pablo and Tanaka, Yukio and Nagaosa, Naoto},
  journal = {Phys. Rev. Lett.},
  volume = {131},
  issue = {9},
  pages = {096001},
  numpages = {7},
  year = {2023},
  month = {Aug},
  publisher = {American Physical Society},
  doi = {10.1103/PhysRevLett.131.096001},
  url = {https://link.aps.org/doi/10.1103/PhysRevLett.131.096001}
}

@article{Lin2022,
  author  = {Lin, Jianxiong and Siriviboon, Phumrojana and Scammell, Harley D. and others},
  title   = {Zero-field superconducting diode effect in small-twist-angle trilayer graphene},
  journal = {\href{https://doi.org/10.1038/s41567-022-01700-1}{Nat. Phys. \textbf{18}, 1221--1227 (2022)}},
}

@article{PhysRevX.12.040002,
  title = {Editorial: Altermagnetism---A New Punch Line of Fundamental Magnetism},
  author = {Mazin, Igor},
  collaboration = {The PRX Editors},
  journal = {Phys. Rev. X},
  volume = {12},
  issue = {4},
  pages = {040002},
  numpages = {3},
  year = {2022},
  month = {Dec},
  publisher = {American Physical Society},
  doi = {10.1103/PhysRevX.12.040002},
  url = {https://link.aps.org/doi/10.1103/PhysRevX.12.040002}
}

@article{PhysRevX.12.040501,
  title = {Emerging Research Landscape of Altermagnetism},
  author = {\ifmmode \check{S}\else \v{S}\fi{}mejkal, Libor and Sinova, Jairo and Jungwirth, Tomas},
  journal = {Phys. Rev. X},
  volume = {12},
  issue = {4},
  pages = {040501},
  numpages = {27},
  year = {2022},
  month = {Dec},
  publisher = {American Physical Society},
  doi = {10.1103/PhysRevX.12.040501},
  url = {https://link.aps.org/doi/10.1103/PhysRevX.12.040501}
}

@misc{hellenes2024pwavemagnets,
      title={P-wave magnets}, 
      author={Anna Birk Hellenes and Tomáš Jungwirth and Rodrigo Jaeschke-Ubiergo and Atasi Chakraborty and Jairo Sinova and Libor Šmejkal},
      year={2024},
      eprint={2309.01607},
      archivePrefix={arXiv},
      primaryClass={cond-mat.mes-hall}, 
}

@article{PhysRevX.12.031042,
  title = {Beyond Conventional Ferromagnetism and Antiferromagnetism: A Phase with Nonrelativistic Spin and Crystal Rotation Symmetry},
  author = {\ifmmode \check{S}\else \v{S}\fi{}mejkal, Libor and Sinova, Jairo and Jungwirth, Tomas},
  journal = {Phys. Rev. X},
  volume = {12},
  issue = {3},
  pages = {031042},
  numpages = {16},
  year = {2022},
  month = {Sep},
  publisher = {American Physical Society},
  doi = {10.1103/PhysRevX.12.031042},
  url = {https://link.aps.org/doi/10.1103/PhysRevX.12.031042}
}

@article{PhysRevX.12.011028,
  title = {Giant and Tunneling Magnetoresistance in Unconventional Collinear Antiferromagnets with Nonrelativistic Spin-Momentum Coupling},
  author = {\ifmmode \check{S}\else \v{S}\fi{}mejkal, Libor and Hellenes, Anna Birk and Gonz\'alez-Hern\'andez, Rafael and Sinova, Jairo and Jungwirth, Tomas},
  journal = {Phys. Rev. X},
  volume = {12},
  issue = {1},
  pages = {011028},
  numpages = {11},
  year = {2022},
  month = {Feb},
  publisher = {American Physical Society},
  doi = {10.1103/PhysRevX.12.011028},
  url = {https://link.aps.org/doi/10.1103/PhysRevX.12.011028}
}

@misc{debnath2024,
      title={Field-free Josephson diode effect in interacting chiral quantum dot junctions}, 
      author={Debika Debnath and Paramita Dutta},
      year={2024},
      eprint={2411.18325},
      archivePrefix={arXiv},
      primaryClass={cond-mat.supr-con}, 
}

@article{Wu2022,
  author  = {Wu, Heng and Wang, Yaojia and Xu, Yuanfeng and Sivakumar, Pranava K. and Pasco, Chris and Filippozzi, Ulderico and Parkin, Stuart S. P. and Zeng, Yu-Jia and McQueen, Tyrel and Ali, Mazhar N.},
  title   = {The field-free josephson diode in a van der waals heterostructure},
  journal = {\href{https://doi.org/10.1038/s41586-022-04504-8}{Nature \textbf{604}, 653--656 (2022)}},
}

@article{PhysRevLett.129.267702,
  title = {Josephson Diode Effect in Supercurrent Interferometers},
  author = {Souto, Rub\'en Seoane and Leijnse, Martin and Schrade, Constantin},
  journal = {Phys. Rev. Lett.},
  volume = {129},
  issue = {26},
  pages = {267702},
  numpages = {6},
  year = {2022},
  month = {Dec},
  publisher = {American Physical Society},
  doi = {10.1103/PhysRevLett.129.267702},
  url = {https://link.aps.org/doi/10.1103/PhysRevLett.129.267702}
}

@article{PhysRevB.110.014518,
  title = {Field-free Josephson diode effect in altermagnet/normal metal/altermagnet junctions},
  author = {Cheng, Qiang and Mao, Yue and Sun, Qing-Feng},
  journal = {Phys. Rev. B},
  volume = {110},
  issue = {1},
  pages = {014518},
  numpages = {12},
  year = {2024},
  month = {Jul},
  publisher = {American Physical Society},
  doi = {10.1103/PhysRevB.110.014518},
  url = {https://link.aps.org/doi/10.1103/PhysRevB.110.014518}
}

@misc{sharma2026pwavemagnetdrivenfieldfree,
      title={$p$-wave magnet driven field-free Josephson diode effect}, 
      author={Lovy Sharma and Bimal Ghimire and Manisha Thakurathi},
      year={2026},
      eprint={2602.16677},
      archivePrefix={arXiv},
      primaryClass={cond-mat.supr-con}, 
}

@misc{salehi2025transversespinsupercurrentpwave,
      title={Transverse Spin Supercurrent at p-wave magnetic Josephson Junctions}, 
      author={Morteza Salehi},
      year={2025},
      eprint={2507.11397},
      archivePrefix={arXiv},
      primaryClass={cond-mat.supr-con}, 
}

@article{Zeng_2025,
   title={Tunneling spin Hall effect induced by unconventional $p$-wave magnetism},
   volume={112},
   ISSN={2469-9969},
   url={http://dx.doi.org/10.1103/twhh-gfyc},
   DOI={10.1103/twhh-gfyc},
   number={14},
   journal={Physical Review B},
   publisher={American Physical Society (APS)},
   author={Zeng, W.},
   year={2025},
   month=Oct }

@article{Baumgartner2022,
  author  = {Baumgartner, Christian and Fuchs, Lorenz and Costa, Andreas and Reinhardt, Maximilian and Paradiso, Nicola and Strunk, Christoph and Swierczynski, Dariusz and Grifoni, Milena and Strunk, Christoph},
  title   = {Supercurrent rectification and magnetochiral effects in symmetric Josephson junctions},
  journal = {\href{https://doi.org/10.1038/s41565-021-01009-9}{Nat. Nanotechnol. \textbf{17}, 39--44 (2022)}},
}

@article{PhysRevLett.131.196301,
  title = {Phase Asymmetry of Andreev Spectra from Cooper-Pair Momentum},
  author = {Banerjee, Abhishek and Geier, Max and Rahman, Md Ahnaf and Thomas, Candice and Wang, Tian and Manfra, Michael J. and Flensberg, Karsten and Marcus, Charles M.},
  journal = {Phys. Rev. Lett.},
  volume = {131},
  issue = {19},
  pages = {196301},
  numpages = {6},
  year = {2023},
  month = {Nov},
  publisher = {American Physical Society},
  doi = {10.1103/PhysRevLett.131.196301},
  url = {https://link.aps.org/doi/10.1103/PhysRevLett.131.196301}
}

@article{PhysRevX.12.041013,
  title = {General Theory of Josephson Diodes},
  author = {Zhang, Yi and Gu, Yuhao and Li, Pengfei and Hu, Jiangping and Jiang, Kun},
  journal = {Phys. Rev. X},
  volume = {12},
  issue = {4},
  pages = {041013},
  numpages = {11},
  year = {2022},
  month = {Nov},
  publisher = {American Physical Society},
  doi = {10.1103/PhysRevX.12.041013},
  url = {https://link.aps.org/doi/10.1103/PhysRevX.12.041013}
}

@misc{zhao2026spinpolarizedjosephsoncurrentinduced,
      title={Spin-polarized Josephson current induced by inhomogeneous altermagnetic interlayers}, 
      author={Wenjun Zhao and Yuri Fukaya and Pablo Burset and Jorge Cayao and Yukio Tanaka and Bo Lu},
      year={2026},
      eprint={2605.02140},
      archivePrefix={arXiv},
      primaryClass={cond-mat.supr-con}, 
}

@misc{patra2026floquetmajoranaflatbands,
      title={Floquet Majorana flat bands and emergent Cooper pair symmetries in $p-$wave magnet$-$superconductor heterostructure}, 
      author={Subhendu Kumar Patra and Gaurab Kumar Dash and Manisha Thakurathi},
      year={2026},
      eprint={2606.31550},
      archivePrefix={arXiv},
      primaryClass={cond-mat.mes-hall}, 
}

@article{PhysRevLett.133.236703,
  title = {Minimal Models and Transport Properties of Unconventional $p$-Wave Magnets},
  author = {Brekke, Bj\o{}rnulf and Sukhachov, Pavlo and Giil, Hans Gl\o{}ckner and Brataas, Arne and Linder, Jacob},
  journal = {Phys. Rev. Lett.},
  volume = {133},
  issue = {23},
  pages = {236703},
  numpages = {9},
  year = {2024},
  month = {Dec},
  publisher = {American Physical Society},
  doi = {10.1103/PhysRevLett.133.236703},
  url = {https://link.aps.org/doi/10.1103/PhysRevLett.133.236703}
}

@article{Zhang2020,
  title = {Nonreciprocal superconducting NbSe$_2$ antenna},
  ISSN = {2041-1723},
  number = {1},
  journal = {\href{https://doi.org/10.1038/s41467-020-19459-5}{Nat. Commun. \textbf{11}, 5634 (2020)}},
  publisher = {Springer Science and Business Media LLC},
  author = {Zhang,  Enze and Xu,  Xian and Zou,  Yi-Chao and Ai,  Linfeng and Dong,  Xiang and Huang,  Ce and Leng,  Pengliang and Liu,  Shanshan and Zhang,  Yuda and Jia,  Zehao and Peng,  Xinyue and Zhao,  Minhao and Yang,  Yunkun and Li,  Zihan and Guo,  Hangwen and Haigh,  Sarah J. and Nagaosa,  Naoto and Shen,  Jian and Xiu,  Faxian},
  month = nov
}

@article{Pal2022,
  author  = {Pal, Banabir and Chakraborty, Anirban and Sivakumar, P. K. and Davydova, Margarita and Gopi, A. K. and Pandeya, A. K. and Krieger, J. A. and Zhang, Yang and Date, Mihir and Ju, Sailong and Yuan, Noah and Schr{\"o}ter, N. B. M. and Fu, Liang and Parkin, S. S. P.},
  title   = {Josephson diode effect from Cooper pair momentum in a topological semimetal},
  journal = {\href{https://doi.org/10.1038/s41567-022-01699-5}{Nat. Phys. \textbf{18}, 1228--1233 (2022)}},
}

@misc{pal2026emergentsuperconductingphasesunconventional,
      title={Emergent superconducting phases in unconventional $p$-wave magnets: Topological superconductivity, Bogoliubov Fermi surfaces and superconducting diode effect}, 
      author={Amartya Pal and Paramita Dutta and Arijit Saha},
      year={2026},
      eprint={2603.03221},
      archivePrefix={arXiv},
      primaryClass={cond-mat.supr-con} 
}

@article{g4ry-j1xy,
  title = {Topological superconductivity and superconducting diode effect mediated via unconventional magnet and Ising spin-orbit coupling},
  author = {Pal, Amartya and Mondal, Debashish and Nag, Tanay and Saha, Arijit},
  journal = {Phys. Rev. B},
  volume = {113},
  issue = {19},
  pages = {195409},
  numpages = {9},
  year = {2026},
  month = {May},
  publisher = {American Physical Society},
  doi = {10.1103/g4ry-j1xy},
  url = {https://link.aps.org/doi/10.1103/g4ry-j1xy}
}
\end{document}